\documentclass[twocolumn]{aastex63}
\usepackage{graphics,graphicx}
\hypersetup{urlcolor=blue}
\usepackage{natbib}
\usepackage[outdir=./]{epstopdf}
\usepackage{textcomp,gensymb}
\usepackage{xfrac}
\usepackage{xcolor}
\usepackage{multirow}
\usepackage{longtable}
\usepackage{booktabs}
\newif\ifrevs
\ifrevs \newenvironment{redblock}{\begingroup\color{red}}{\endgroup}
\else   \newenvironment{redblock}{}{}
\fi

\newcommand{\kepler}{{\it Kepler}}

\newcommand{\rchisq}{$\chi^2_{\nu}$}

\newcommand{\gaia}{{\textit Gaia}}

\usepackage[outdir=./]{epstopdf}
\usepackage{ulem}
\usepackage{mathrsfs}
\usepackage{fontawesome}
\usepackage{comment}
\usepackage[version=4]{mhchem}

\newcommand{\tess}{\textit{TESS}}
\newcommand{\ktwo}{{\textit K2}}

\newcommand{\jwst}{\textit{JWST}}

\pdfoutput=1

\shorttitle{TTVs of young planets} 
\shortauthors{Lopez Murillo et al.}

\begin{document}

\title{Searching for Transit Timing Variations in young transiting systems}

\correspondingauthor{Ana Isabel Lopez Murillo}
\email{isaana@email.unc.edu} 

\author[0009-0006-9572-1733]{Ana Isabel Lopez Murillo}
\altaffiliation{UNC Chancellor’s Science Scholar}
\altaffiliation{NC Space Grant Graduate Research Fellow}
\affiliation{Department of Physics and Astronomy, The University of North Carolina at Chapel Hill, Chapel Hill, NC 27599, USA}

\author[0000-0003-3654-1602]{Andrew W. Mann}
\affiliation{Department of Physics and Astronomy, The University of North Carolina at Chapel Hill, Chapel Hill, NC 27599, USA}

\author[0000-0002-8399-472X]{Madyson G. Barber}
\altaffiliation{NSF Graduate Research Fellow}
\affiliation{Department of Physics and Astronomy, The University of North Carolina at Chapel Hill, Chapel Hill, NC 27599, USA} 

\author[0000-0001-7246-5438]{Andrew Vanderburg}
\affiliation{Department of Physics and Kavli Institute for Astrophysics and Space Research, Massachusetts Institute of Technology, Cambridge, MA 02139, USA}

\author[0000-0001-5729-6576]{Pa Chia Thao}
\altaffiliation{NSF Graduate Research Fellow}
\altaffiliation{Jack Kent Cooke Foundation Graduate Scholar}
\affiliation{Department of Physics and Astronomy, The University of North Carolina at Chapel Hill, Chapel Hill, NC 27599, USA} 

\author[0000-0001-6037-2971]{Andrew W. Boyle}
\affiliation{Department of Physics and Astronomy, The University of North Carolina at Chapel Hill, Chapel Hill, NC 27599, USA}

\begin{abstract} 
The discovery of young ($<$800\,Myr) transiting planets has provided a new avenue to explore how planets form and evolve over their lifetimes. Mass measurements for these planets would be invaluable, but radial velocity surveys of young systems are often overwhelmed by stellar activity. Transit timing variations (TTVs) offer an alternative route to measure masses that are less impacted by signals from the host star. Here we search for candidate TTVs in a sample of 39 young systems hosting 53 transiting planets using data from \kepler{}, \ktwo{}, and \tess. We recover previously reported TTVs for 11 planets, including those in V1298 Tau, TOI-2076, Kepler-51, and TOI-1227, and identify new candidate TTVs for four planets (DS Tuc Ab, HD\,63433b, K2-101b, and Kepler-1643b). In total, $28.3\pm6.2$\% of young planets in our sample show evidence of TTVs, which is higher than the rate from \kepler\ on mostly older systems ($7.3\pm0.6\%)$. Accounting for differences in data coverage and quality between \kepler\ and \tess\ only increases this difference ($>4\sigma$), although differences in methodology make a totally fair comparison challenging. We show that spots have a weak-to-negligible impact on our results, and similarly cannot explain the higher TTV fraction. Longer-term monitoring will be required to validate these TTVs as planetary in nature and confirm the high TTV rate. While the candidate TTV signals detected here are sparsely sampled, our work provides a clear priority list for additional ground-based observations, and for multi-planet TTVs, attempt to measure the masses and eccentricities of these planets.
\end{abstract}


\section{Introduction}\label{sec:intro}
Studying planets across a wide range of ages offers a unique window into how planets form and evolve. Planets with ages $\ll$1\,Gyr are the most valuable, as planets evolve faster while they are young. However, the majority of discovered planetary systems are either old ($>$1\,Gyr) or have poorly constrained ages \citep{Berger2020, Bouma2024}. Fortunately, the \ktwo\ and \tess\ missions surveyed a number of nearby young clusters and associations, leading to the discovery of a large population of young (3-800\,Myr), transiting planets with precise ages derived from their parent population \citep[e.g.,][]{THYMEIII,Barber2022, THYMEVII}.

Theoretical models predict that young ($\ll$ 1\, Gyr) planets will be larger than their older counterparts due to the retained heat from their formation that causes their atmosphere to expand \citep{Lopez2012}. However, there are a range of physical drivers, and it remains unclear which is the most important. Low-mass, H/He-rich planets can shrink by a factor of $\simeq$2 in radius between 50 Myr and 500 Myr due to thermal contraction \citep{Lopez2014}. Young planets may also appear larger because of the presence of photochemical hazes that render their atmospheres opaque \citep{Gao_hazes}. Alternatively, the low surface gravity and high irradiation of most young planets make them vulnerable to atmospheric mass-loss through photo-evaporation \citep{Murray-Clay2009}, driving the planets to smaller radii as material of low mean molecular weight is lost. All of these processes operate on a different duration, and studying young planets will provide a timescale for changes.

Masses for a sample of the young planets are essential to break these degeneracies and provide a more quantitative test of planetary evolution models. A young planet may appear to have a similar radius to an older planet but still be significantly inflated if it is less massive. Without accurate mass estimates, it is difficult to interpret the detection of features in transmission spectra, as there are strong degeneracies between surface gravity and composition \citep{Batalha2017}. Further, masses are essential to perform a fair comparison between old and young planetary systems given the complex relation between period, radius, and migration history.

Radial velocity work in young systems has proven to be challenging. The initial mass detection of 30\,Myr V1298 Tau\,e claimed in \citet{SuarezMascareno2021} found an orbital period 4$\sigma$ discrepant with the follow-up transit data in \cite{Feinstein2022}, and in conflict with more extensive RV data \citep{Sikora2023, Blunt2023}. \citet{Zakhozhay2022} found a mass of 8$\pm1M_J$ for the 15\,Myr Jupiter-radius planet HD 114082\,b, but even cold-start models predict a mass below $1M_J$ at this age \citep{Spiegel2012}, and the the report of a second {\it larger} planet in the system \citep{EXOFOP_TESS} invalidates their analysis. A number of other systems have been the subject of extensive monitoring campaigns and only yielded upper limits on mass \citep[e.g.,][]{Benatti2021, Damasso2023}. Similarly concerning is that the youngest planets discovered through radial velocities are not consistently recovered \citep[e.g.,][]{Donati2020, Damasso2020}. 

It is possible to estimate the mass of a planet through its transmission spectrum by taking advantage of the relation between the planet's scale height and its surface gravity \citep{deWit2013}. Applying this method to small planets is challenging due to degeneracies between metallicity, cloud cover, stellar surface inhomogeneities, and surface gravity \citep{Batalha2017}. The method has worked, however, on young gas giants \citep{Thao2024_featherweight, Barat2024}. Such systems tend to have lower surface gravity than their older counterparts, making it easier to break the degeneracies. This method may still fail for the smallest planets, and measuring masses via transmission spectroscopy for a large sample of young planets would require a prohibitive amount of telescope time. 

An alternative route for precise masses of young planets is Transit Timing Variations (TTVs) -- changes in the planet's transit time due to the gravitational influence of another object. TTVs have provided masses for small planets \citep{Jontof-Hutter2015} and a limited set of young stars \citep[e.g.,][]{Masuda2014}. Importantly, young planets are found near mean-motion resonances (MMR) at a much higher rate than for older planets \citep[e.g., 86$\pm$13\% of multis $<$100\,Myr;][]{Dai2024}. Since TTV signals are much larger for planets near MMR, we expect a similarly high fraction of young multi-transiting systems to show TTVs.

Often a detected timing variation is insufficient to constrain the full system parameters. Binaries and Roemer delay can cause TTVs without any planet-planet interaction. Even if the signal is from another planet, it is not always possible to break degeneracies between eccentricity and mass \citep{becker2015}. However, such TTVs may still be useful for setting limits on eccentricity and mass, detect additional planets in the system, and motivate future ground-based studies to fully map out the TTV signal. Indeed, the confirmation of HIP~67522\,c \citep{Barber_HIPc} was partially motivated by the detection of TTVs in the inner planet \citep{Thao2024_featherweight}.

Here we search for signs of TTVs in young transiting planets with \kepler, \ktwo, and/or {\it TESS} data. In Section~\ref{sec:target} we list the young systems of interest. We describe the light curve data in Section~\ref{sec:obs}. We then detail our method for fitting individual transits and searching for TTVs in Section~\ref{sec:methods}. In Section~\ref{sec:results} we overview which systems show significant TTVs, assess the role of spots on the TTV signal, and compare our TTV fraction to that of older systems from \kepler. We conclude in Section~\ref{sec:conclusion} with a discussion of which TTVs show the best promise for mass detections in the immediate future.

\section{Target Selection }\label{sec:target}

We initially selected all transiting planets with ages $<$800\,Myr and age uncertainties $<$100\,Myr. The former cut is roughly Hyades age or younger, where stellar activity can make radial velocities more challenging \citep{Cochran2002, Paulson2004, Brems2019}. The latter is meant to remove targets with highly uncertain ages, many of which are likely to be $>$1\,Gyr due to a strong bias in favor of older systems in the transiting planet sample. We drew this list from a mix of the NASA Exoplanet Archive \citep{Christiansen2025} and a search of recent literature \citep[e.g.,][]{THYMEVII}. 

The sample was finalized in September 2024. Planets discovered after this date were not included. The exception was IRAS 04125+2902\,b, which already included an (identical) TTV analysis to that performed here as part of the discovery paper \citep{Barber_IRAS}. 

From this list, we removed four systems. HD\,114082 \citep{Zakhozhay2022} and HD\,283869 \citep{Vanderburg2018} are both single-transiting systems, for which TTV detections will not be possible. Kepler-411 has a reported age of 200\,Myr  \citep{Sun2019}, but the host star shows no detectable lithium \citep{Berger2018} and the 10.5\,day rotation period and color puts it in the `stalling' regime \citep{Curtis_stall}. In combination, this suggests an age of at least 600\,Myr. TOI-560 has the same problem: no lithium and a rotation period consistent with 0.1--1.4\,Gyr. While it is possible these meet our age criteria, they would not meet our age uncertainty criteria, so we opted to remove them.

We additionally cut planets with $<3$ transits in their combined datasets. This removes seven planets, all of which are outer-planets in multi-transiting systems (AU\,Mic\,c, K2-136\,d, K2-233\,d, TOI-2076\,c, TOI-2076\,d, TOI-964\,c, V1298\,Tau\,e). Notably, this removes many planets with known TTVs \citep[e.g.][]{Osborn2022,Wittrock2022}. We additionally removed TOI-1860 due to the complications with the fitting procedure, as outlined below.

This left 53 planets orbiting 39 stars. The majority have robust ages derived from membership to a young association, but many have ages derived from a combination of gyrochronology and similar indicators of age (e.g., lithium). We provide a full list in Table~\ref{tab:obs}.

\begin{longrotatetable}
\begin{deluxetable*}{lccccrrrrrr} 
\tabletypesize{\footnotesize} 
\tablecaption{Overview of Input Light Curves}
\tablewidth{0pt}
\tablehead{
\colhead{System} & 
\colhead{planet} &
\colhead{Transiting} &
\colhead{TESS sectors} &
\colhead{Other input data} &
\colhead{Time Span} &
\colhead{$N_{\rm{transits}}$} &
\colhead{Significance } &
\colhead{Parameters\tablenotemark{$\alpha$}} \\
\colhead{} & 
\colhead{} & 
\colhead{planets} & 
\colhead{} & 
\colhead{} &
\colhead{(years)} &
\colhead{} &
\colhead{$\chi^2 / d.o.f$} &
}
\startdata          
    AU\,Mic         & b & 2 & 1, 27                         & - & 0.15 & 5 & 5.62 & \citenum{Wittrock2023} \\
    DS\,Tuc A       & b & 1 & 1, 27, 28, 67, 68             & - & 0.37 & 15 & 2.88 & \citenum{Newton2019} \\
    HD\,110082      & b & 1 & 12, 13, 27, 65, 66, 67        & - & 0.44 & 16 & 1.20 & \citenum{THYMEV} \\
    HD\,178085      & b & 1 & 27, 66, 67                    & - & 0.22 & 6 & 2.78 & \citenum{Zhou2022, Mantovan2022} \\
    HD\,63433       & b & 4 & 20, 44, 45, 46, 47, 72        & - & 0.44 & 18 & 2.41 & \citenum{Polanski2024, Capistrant2024} \\
                    & c & - & 20, 44, 45, 46, 47, 72        & - & 0.44 & 8 & 1.06 & \citenum{Polanski2024, Capistrant2024} \\
                    & d & - & 20, 44, 45, 46, 47, 72        & - & 2.3 & 29 & 0.72 & \citenum{Polanski2024, Capistrant2024} \\
    HIP\,67522      & b & 2 & 11, 38, 64                    & - & 0.22 & 10 & 0.53 & \citenum{Barber_HIPc} \\ 
                    & c & - & 11, 38, 64                    & - & 0.22 & 5 & 0.76 & \citenum{Barber_HIPc} \\
    IRAS04125+2902  & b & 1 & 19, 43, 44, 59, 70, 71        & - & 0.44 & 16 & 0.71 & \citenum{Barber_IRAS} \\
    K2\,100         & b & 1 & 44, 45, 46, 72                & K2 Campaign 5, 18 & 0.8 & 129 & 0.46 & \citenum{Barragan2019, Stefansson2018} \\
    K2\,101         & b & 1 & -                             & K2 Campaign 5, 16, 18 & 0.75 & 13 & 5.07 & \citenum{CastroGonzalez2022, Mann2017a} \\
    K2\,102         & b & 1 & -                             & K2 Campaign 5, 16, 18 & 0.75 & 20 & 1.49 & \citenum{Mann2017a} \\
    K2\,103         & b & 1 & -                             & K2 Campaign 5, 16, 18 & 0.75 & 10 & 1.57 & \citenum{CastroGonzalez2022, Dressing2017, Mann2017a} \\
    K2\,136         & b & 3 & -                             & K2 Campaign 13        & 0.25 & 9 & 0.97 & \citenum{Mayo2023} \\
                    & c & - & 43, 44, 70, 71                & K2 Campaign 13        & 0.47 & 10 & 3.07 & \citenum{Mayo2023, Kokori2023} \\
    K2\,233         & b & 2 & -                             & K2 Campaign 15        & 0.25 & 35 & 0.29 & \citenum{Lillo-Box2020}\\
                    & c & - & -                             & K2 Campaign 15        & 0.25 & 11 & 0.30 & \citenum{Lillo-Box2020}\\
    K2\,25          & b & 1 & 44, 70, 71                    & -                     & 0.22 & 20 & 0.38 & \citenum{Mann2016a,Stefansson2020}\\
    K2\,264         & b & 2 & -                             & K2 Campaign 16        & 0.25 & 13 & 0.71 & \citenum{Livingston2019} \\
                    & c & - & -                             & K2 Campaign 16        & 0.25 & 4 & 2.46 &  \citenum{Livingston2019} \\
    K2\,284         & b & 1 & 43, 44, 45, 71                & K2 Campaign 13        & 0.55 & 34 & 0.80 & \citenum{David2018} \\
    K2\,33          & b & 1 & -                             & K2 Campaign 2         & 0.25 & 14 & 0.55 & \citenum{Mann2016b} \\
    K2\,95          & b & 1 & -                             & K2 Campaign 5, 18     & 0.5  & 13  & 1.31 & \citenum{CastroGonzalez2022, Mann2017a}, \\
    KOI\,7913 A     & b & 1 & -                             & Kepler Q1-17          & 4.25 & 55  & 0.29 & \citenum{Bouma2022} \\
    Kepler\,1627 A  & b & 1 & -                             & Kepler Q1-15          & 4    & 146 & 1.50 &  \citenum{Morton2016, KOICatalog}\\
    Kepler\,1643    & b & 1 & 14, 41, 54, 55, 74, 75, 81, 82 & Kepler Q1-17        & 4.84 & 490 & 23.61 &  \citenum{Bouma2022, KOICatalog} \\
    Kepler\,1928    & b & 1 & -                              & Kepler Q1-17        & 4.25 & 73 & 1.26 & \citenum{Barber2022}\\
    Kepler\,51      & b & 3& -                             & Kepler Q1-17          & 4.25 & 30 & 4.84 & \citenum{Masuda2014, Libby-Roberts2020} \\ 
                    & c & -& -                             & Kepler Q1-17          & 4.25 & 15 & 7.65 & \citenum{Masuda2014, Libby-Roberts2020}\\ 
                    & d & -& -                             & Kepler Q1-17          & 4.25 & 10 & 346.20 & \citenum{Masuda2014, Libby-Roberts2020} \\ 
    Kepler\,63      & b & 1& 14, 15, 40, 41, 53, 54, 55, 74, 75, 80, 81, 82        & Kepler Q2-17 & 4.85 & 170 & 0.94 & \citenum{SanchisOjeda2013} \\
    Kepler\,970     & b & 1 & -                                                    & Kepler Q1-17 & 4.25 & 80 & 2.42 & \citenum{Barber2022} \\
    Qatar\,3        & b & 1 & 57                                                   & - & 0.07 & 10 & 0.51 & \citenum{Alsubai2017, Kokori2023} \\
    Qatar\,4        & b & 1 & 57                                                   & - & 0.07 & 14 & 0.55 & \citenum{Alsubai2017, Kokori2023} \\
    Qatar\,5        & b & 1 & 14, 17, 57                                           & - & 0.22 & 15 & 1.08 & \citenum{Alsubai2017, Kokori2023} \\
    TOI\,1097       & b & 2 & 38, 39, 65                                           & - & 0.22 & 8 & 2.05 & \citenum{Wood2023} \\
                    & c & - & 38, 39                                               & - & 0.15 & 4 & 2.82 & \citenum{Wood2023} \\
    TOI\,1224       & b & 2 & 1,13,27,28,39,66-68                                  & - & 0.52 & 47 & 0.91 & \citenum{Thao2024a} \\
                    & c & - & 1,13,27,28,39,66-68                                  & - & 0.52 & 12 & 4.50 & \citenum{Thao2024a} \\
    TOI\,1227       & b & 1 & 11, 12,38,64,65                                      & - & 1.33 & 5 & 12.35 & \citenum{THYMEVI, Almenara2024} \\
    TOI\,1807       & b & 1 & 22, 23, 49, 76                                       & - & 0.30 & 142 & 1.21 & \citenum{Polanski2024, Nardiello2022} \\
    TOI\,2048       & b & 1 & 16, 23, 24, 50, 51, 77, 78                           & - & 0.52 & 9 & 0.61 & \citenum{THYMEVII} \\
    TOI\,2076       & b & 3 & 16, 23, 50, 77                                       & - &  0.28 & 9 & 4.13 & \citenum{Polanski2024, Kokori2023} \\
    TOI\,251        & b & 1 & 1, 2, 29, 69                                         & - & 0.28 & 13 & 0.40 & \citenum{Zhou2021} \\
    TOI\,451        & b & 3 & 4, 5, 31                                             & - & 0.22 & 39 & 0.99 & \citenum{Newton2021}\\
                    & c & - & 4, 5, 31                                             & - & 0.22 & 8 & 0.96 & \citenum{Newton2021}\\
                    & d & - & 4, 5, 31                                             & - & 0.22 & 5 & 1.63 & \citenum{Newton2021} \\
    TOI\,837        & b & 1 & 10, 11, 37, 38, 63, 64                               & - & 0.44 & 21 & 0.59 & \citenum{Barragan2024, Bouma2020} \\
    TOI\,942        & b & 2 & 32                                                   & - & 0.07 & 6 & 1.87 & \citenum{Carleo2021, Wirth2021}\\
    V1298\,Tau      & b & 4 & 43, 44                               & K2 Campaign 4 & 0.75 & 5 & 5.50 & \citenum{David:2019aa, FeinsteinV1298Tau2022} \\
                    & c & - & 43, 44                                           & - & 0.22 & 6  & 2.84 & \citenum{David:2019aa, FeinsteinV1298Tau2022} \\
                    & d & - & 43, 44                               & K2 Campaign 4 & 0.75 & 9 & 3.18 & \citenum{David:2019aa, FeinsteinV1298Tau2022} \\
\enddata
\tablenotetext{\alpha}{ 
    \scriptsize 
        \citenum{Wittrock2023} = \citet{Wittrock2023}; 
        \citenum{Newton2019} = \citet{Newton2019}; 
        \citenum{THYMEV} = \citet{THYMEV}; 
        \citenum{Zhou2022} = \citet{Zhou2022}; 
        \citenum{Mantovan2022} = \citet{Mantovan2022}; 
        \citenum{Polanski2024} = \citet{Polanski2024}; 
        \citenum{Capistrant2024} = \citet{Capistrant2024}; 
        \citenum{Barber_HIPc} = \citet{Barber_HIPc};
        \citenum{Barber_IRAS} = \citet{Barber_IRAS}; 
        \citenum{Barragan2019} = \citet{Barragan2019}; 
        \citenum{Stefansson2018} = \citet{Stefansson2018}; 
        \citenum{CastroGonzalez2022} = \citet{CastroGonzalez2022}; 
        \citenum{Mann2017a} = \citet{Mann2017a}; 
        \citenum{Bouma2022} = \citet{Bouma2022} 
        \citenum{Morton2016} = \citet{Morton2016}; 
        \citenum{KOICatalog} = \citet{KOICatalog}; 
        \citenum{Barber2022} = \citet{Barber2022};  
        \citenum{Masuda2014} = \citet{Masuda2014};
        \citenum{Libby-Roberts2020} = \citet{Libby-Roberts2020};
        \citenum{SanchisOjeda2013} = \citet{SanchisOjeda2013};
        \citenum{Alsubai2017} = \citet{Alsubai2017};
        \citenum{Kokori2023} = \citet{Kokori2023};
        \citenum{Wood2023} = \citet{Wood2023};
        \citenum{Thao2024a} = \citet{Thao2024a};
        \citenum{THYMEVI} = \citet{THYMEVI};
        \citenum{Almenara2024} = \citet{Almenara2024};
        \citenum{Nardiello2022} = \citet{Nardiello2022};
        \citenum{Giacalone2022} = \citet{Giacalone2022};
        \citenum{THYMEVII} = \citet{THYMEVII};
        \citenum{Zhou2021} = \citet{Zhou2021};
        \citenum{Newton2021} = \citet{Newton2021};
        \citenum{Barragan2024} = \citet{Barragan2024};
        \citenum{Bouma2020} = \citet{Bouma2020};
        \citenum{Carleo2021} = \citet{Carleo2021};
        \citenum{Wirth2021} = \citet{Wirth2021};
        \citenum{David:2019aa} = \citet{David:2019aa};
        \citenum{FeinsteinV1298Tau2022} = \citet{FeinsteinV1298Tau2022};
}
\label{tab:obs}
\end{deluxetable*}
\end{longrotatetable}

\begin{deluxetable*}{lccc}
\centering
\tablecaption{Corrected Planetary Parameters \label{tab:periods}}
\tablecolumns{4}
\tablewidth{0pt}
\tablehead{
\colhead{System} & 
\colhead{Planet} & 
\colhead{TTV Corrected Period (days)} &
\colhead{TTV Corrected $T_0$ (days)}
}
\startdata
    \hline
    AU Mic & b &    $8.4629993$        & - \\
    DS Tuc A & b &  $8.1382616$        & - \\
    K2-264 & c &    -        & $2458116.4187$ \\
    Kepler 63 & b & -                  & $2455010.844$ \\
    Kepler-51 & d & -                  & $2455045.0127$ \\
    Kepler-1627 A & b & $7.2028330$   & - \\
    Qatar 5 & b &   $2.87929817$      & - \\
    TOI-1097 & c &  -                 & $2456441.05804$ \\
    TOI-1224 & c &  $17.945444$       & - \\
    TOI-2076 & b &  $10.355106$       & - \\
    \\
    \hline
\enddata 
\end{deluxetable*}

\section{Light Curves}\label{sec:obs}

We used light curves from \kepler, \ktwo, and \tess, where possible. While nearly all targets have \tess\ data, many planets detected in \ktwo\ or \kepler\ data orbit stars that are too faint or land in fields too crowded to make full use of the \tess\ data. Further, for our TTV analysis, it is important that each individual transit be detectable to extract a usable transit time. For some smaller planets, the signal is only significant over the full curve, particularly in the \tess\ data. We used two-minute (120 second) cadence \tess\ data when available, and \tess\ FFI light curves of 10-minute or 200-second cadence for sectors that did not have two-minute data available. Tests showed that using longer-cadence data was insufficient to resolve the time of transit midpoint due to instrument noise, so we opted to use data of 10 minutes or shorter for our analyses. Shorter FFI data was included on a case-by-case basis, and we list what datasets were included for each planet in Table~\ref{tab:obs}.

We have found that, in general, reported errors from SAP and PDCSAP light curves are significantly underestimated. For \ktwo\ and \tess, we are also generating the curves ourselves, so we need a method to compute uncertainties. For all light curves, we estimated the flux uncertainties from the point-to-point scatter in the data itself following \cite{Barber2024}. Specifically, we used the median absolute deviation of the curve after subtracting the data shifted by one point. Past work suggests this is effective over a wide range of curves \citep[e.g.,][]{Mann2016b,Mann2017b}.

\subsection{\tess}

For young active stars, the default \tess\ data processing can fail due to complications separating instrumental and stellar signals \citep[e.g.,][]{Taaki2024}. We follow the systematics correction to the simple aperture photometry (SAP) from the SPOC pipeline \citep{Jenkins2016} outlined in \citet{Vanderburg2019}. This method has dramatically improved the noise properties of the light curves for young stars \citep[e.g.][]{Barber_HIPc, Thao2024a}. 

We confirm the advantage of the modified pipeline by analyzing one system (HIP\,67522\,b and c) with both PDCSAP (Pre-search Data Conditioning SAP) and the \citet{Vanderburg2019} method. While the results were broadly consistent, the per-planet fit was better (better reduced $\chi^2$) using the \citet{Vanderburg2019} light curve extraction.

For \tess\ light curves displaying significant stellar flares, we used \texttt{stella}\footnote{\faGithub \url{https://github.com/afeinstein20/stella}} \citep{Feinstein2020_stella}, a neural network approach to identifying stellar flares. Using the 10 models established in the \citep{Feinsteinstella} training set, we obtained an average flare prediction for each data point. We then fed the \tess\ light curve data in to the network. Flare removal was done on a case-by-case basis, and points with a $>$ 80\% probability of being flares were removed from our sample.

All \tess\ data used in the correction analysis can be found on Mikulski Archive for Space Telescopes (MAST) in the ``TESS `Fast' Light Curves - All Sectors" \citep{TESS_fast}, ``TESS Light Curves - All Sectors" \citep{TESS_LC}, and ``TESS Raw Full Frame Images: All Sectors" \citep{TESS_ffi} repositories.

\subsection{\kepler}

For the seven systems with \kepler\ data, we used the Pre-search Data Conditioning Simple Aperture Photometry \citep[PDCSAP;][]{Smith2012, Stumpe2014} light curve produced by SPOC and available through MAST 
in the ``Kepler LC+SC,  Q0-Q17" repository \citep{kepler_lc}.

In order to ensure data reliability, we only considered data points with \texttt{DQUALITY=0}, indicating that no flags were raised by the SPOC pipeline.

\subsection{\ktwo}

The \ktwo\ mission operated with only two reaction wheels, causing the telescope to shift. During the drift and subsequent thruster fire, a stellar image moves over the detector. Combined with variations in the pixel sensitivity, this generates changes in estimated flux from a given star as a function of centroid position. To correct for this, we started with light curves from the \texttt{K2SFF} pipeline \citep{Vanderburg2014}. The baseline \texttt{K2SFF} pipeline can fail on stars where the variability operates on timescales comparable to the telescope drift because by default, it only modeled variability with timescales significantly longer than 1.5 days. Following prior analyses \citep{Mann2017a, Vanderburg2018, Thao2020}, we used a custom version of \texttt{K2SFF} that models variability with splines with more closely-spaced breakpoints. Once preliminary light curves were generated using this modified pipeline, we refined the systematics correction by performing a simultaneous fit of the transits, variability and the instrumental systematics following \citet{Vanderburg2016ApJS}.

All \ktwo\ data used in the correction analysis can be found on MAST 
in the ``K2 Light Curves (all)" repository \citep{k2_lc}.

\section{Methods}\label{sec:methods}

\subsection{Measuring Individual Transit Times}

A summary of our process is as follows:
\begin{enumerate}
    \item For each planet, we download all available light curve data (see subsections above).
    \item We take initial planet parameters from the NASA Exoplanet Archive as noted in Table \ref{tab:obs}. In some cases, we retrieve parameters directly from the discovery paper.
    \item We fit each transit (one at a time) using a \texttt{BATMAN} model \citep{Kreidberg2015} and a Gaussian Process (GP) to handle stellar variability. For 30-minute cadence data, we initialized the \texttt{BATMAN} model with a supersample factor of 10 to account for long exposures.
    \item We inspect the fits and residuals. If the transit occurs near a data edge or the transit is unusable due to flares, we remove the transit.
    \item We characterize the significance of the TTV initially on reduced $\chi^2$ (\rchisq) of the deviation from a linear ephemeris (O-C) and a BIC comparison between a linear fit and a sine curve of polynomial.
\end{enumerate} 

For each transit fit (step 3), we locked all parameters to the literature value except the transit midpoint ($T_0$) and three GP parameters. For the GP, we use a simple harmonic oscillator (SHO) as implemented in \texttt{Celerite2} \citep{Foreman-Mackey2018,celerite2}. This includes three free parameters: the standard deviation of the process ($\sigma$), the underdamped period of the oscillator ($\rho$), and the quality factor ($Q$). Tests using simpler polynomials yielded poor fits to the data. Other quasi-periodic kernels worked similarly well, although we preferred the SHO kernel because of a long history of success on young stellar systems \citep[e.g.,][]{Thao2023, Wood2023, Barber_IRAS}.

\begin{redblock}
The likelihood is computed inside \texttt{Celerite2} using the residual:
\begin{equation}
\mathbf{r} \equiv \mathbf{y} - m(\mathbf{t}\,|\,\boldsymbol{\theta}), 
\end{equation}
where $m(\mathbf{t}\,|\,\boldsymbol{\theta})$ represents the \texttt{BATMAN} model, $\theta$ is the transit parameters ($T_0$), and $y$ is the flux at time $t$. The other component of the likelihood function is the covariance, which is modeled as:
\begin{equation}\label{eqn:covariance}
\mathbf{C}(\boldsymbol{\phi}, \boldsymbol{\sigma}, \mathbf{s}) \;=\;
\mathbf{K}_{\rm SHO}(\mathbf{t};\,\boldsymbol{\phi})
\;+\; \mathrm{diag}\!\big(\boldsymbol{\sigma}^2\big),
\end{equation}
where ${K}_{\rm{SHO}}$ is the SHO kernel, with hyperparameters $\theta$, and $\sigma$ represents the quoted per-point flux uncertainties. With this, the log-likelihood is:
\begin{equation}
\ln \mathcal{L}(\boldsymbol{\theta},\boldsymbol{\phi},\mathbf{s})
= -\tfrac{1}{2}\,\mathbf{r}^\top \mathbf{C}^{-1} \mathbf{r}
  -\tfrac{1}{2}\,\ln\!\det \mathbf{C}
  -\tfrac{N}{2}\,\ln(2\pi).
\end{equation}

Some fits include an extra scatter term, such that $\sigma_{\mathrm{eff},i}^2 \equiv \sigma_i^2 + s^2$ in Equation~\ref{eqn:covariance} to handle underestimated uncertainties in the data. We opted not to do this, primarily because we already measure our flux uncertainties from the scatter in the data itself (Section~\ref{sec:obs}). As a result, when we tested adding such a parameter on a sample of systems, it was almost always either unbound ($\log{S}\to-\infty$) or negligibly small compared to the formal uncertainties ($<$1\%). The exception was with astrophysical signals (flares) where the fit compensates for its inability to model the data by overestimating $s$ (and hence overestimating uncertainties on the fit parameters). 

\end{redblock}

For each individual transit, we started the fit using the expected linear ephemeris and placed bounds on $T_0$ (initially $\pm72$\,min) to prevent walkers from wandering off the true transit. We sometimes narrowed the window to avoid nearby flares (which the fit sometimes mistakes for an ingress/egress) or if another planetary transits occurs nearby.

The transit fits were done using a Monte Carlo Markov Chain framework with \texttt{emcee} \citep{Foreman-Mackey2013}. All parameters evolved under uniform priors. Every transit ran for at least 15,000 steps with 30 walkers. At this point, we checked that fits ran for at least 50$\times$ the autocorrelation time, as estimated using \texttt{emcee}'s autocorrelation parameter. This is usually sufficient for convergence \citep{Goodman2010}. A small number of fits that did not pass this threshold were allowed to run for up to 40,000 steps (at which point all passed). 
 
\begin{figure*}[ht]
    \centering
    \includegraphics[width=0.90\textwidth]{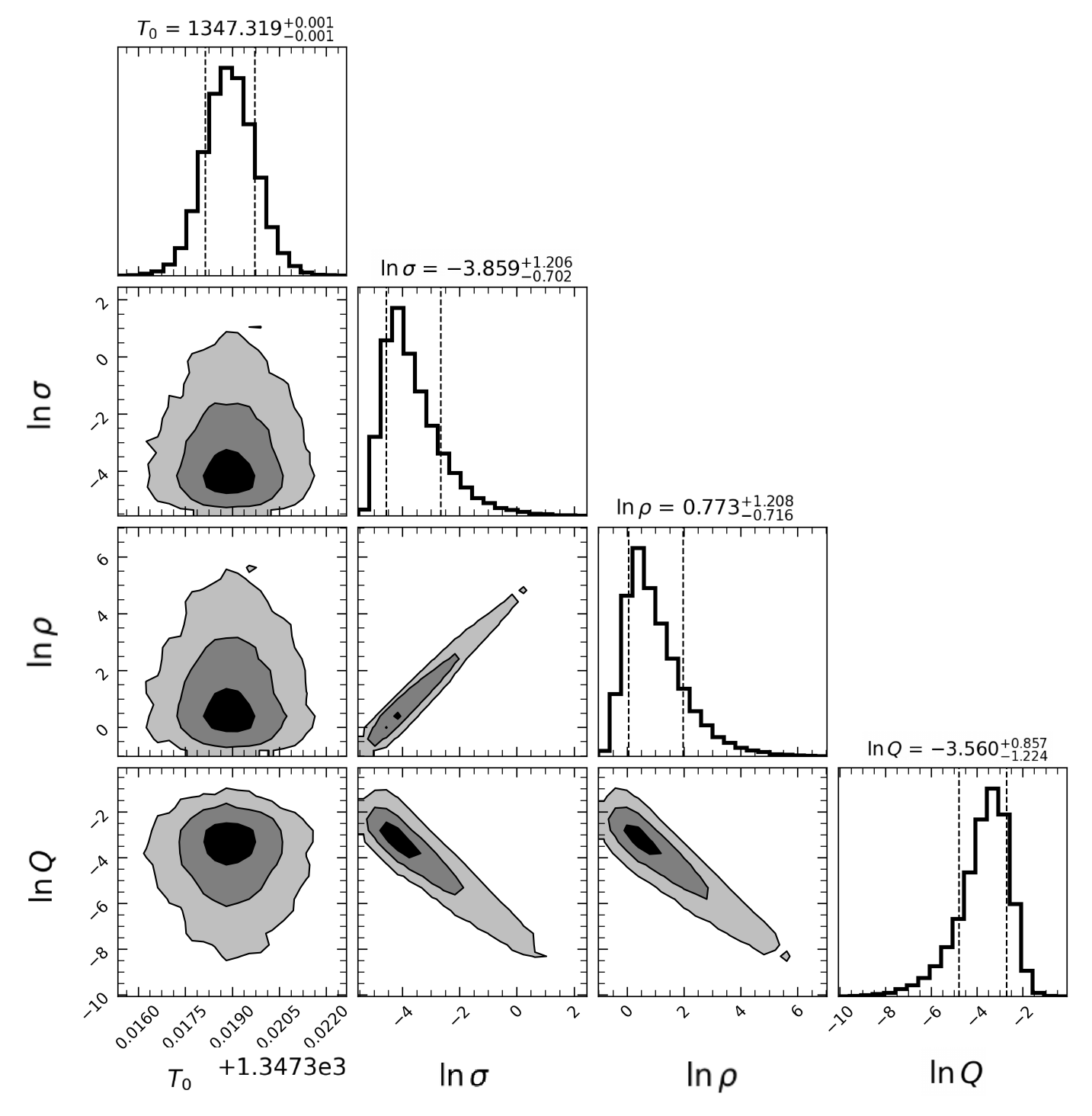}  
    \caption{Example corner plot for the second transit of AU Mic b showing the covariance between (primarily GP) parameters. Other transit parameters are fixed during the fit. The resulting posterior on $T_0$ is Gaussian, which is typical for most transits and planets.
    \label{fig:corner}}
\end{figure*}

We show an example fit posterior in Figure~\ref{fig:corner}. The resulting transit time posteriors were almost universally Gaussian. Those deviating from a Gaussian were those with Gaussian cores with wider tails or slight asymmetries. For simplicity, we opted to convert these to one-dimensional uncertainties when computing the TTV signal. 

A small number ($\lesssim$10 of almost 2000) of transits were removed after visual inspection. These fall under two categories: 1) a data gap/edge where the data quality was lower and/or a large part of the transit was missing or 2) multiple flares that were not captured by \texttt{stella} or modeled out by the GP that the fit confused for ingress/egress. In these cases, the initial fit always favored a transit time off from the expected value; including them would create a bias in favor of TTVs.

Using the literature reported period and $T_0$, we computed the difference between the expected and measured transit times (the O-C). Some systems, including DS Tuc A b, displayed an offset in TTVs from the $y=0$ line or a linear trend due to uncertainties in the reported orbital period or $T_0$ value (i.e., the new data suggests a slightly different orbital period or $T_0$.) We updated these values by fitting a simple straight line fit to correct for this offset, and reported those with significant changes in Table~\ref{tab:periods}.

We compared the resulting O-C points to a flat-line model and computed a reduced $\chi^2$ (\rchisq). This is done after correcting for period and $T_0$ differences (so offset and slope are included). The output \rchisq{} are reported in Table~\ref{tab:obs} for each planet.

In addition to the $\chi^2$ test, we fit each $\mathrm{OC}(t)$ dataset with three functional forms---a linear trend, 
a polynomial of degree $d$, and a sine curve with a constant offset: 
\begin{equation}
\mathrm{OC}(t)=
\begin{cases}
m\,t + b, & \text{(linear)}\\[6pt]
\displaystyle \sum_{p=0}^{d} c_p\,t^{\,p}, & \text{(polynomial of degree $d$)}\\[10pt]
A\,\sin\!\left(\dfrac{2\pi t}{P}+\phi\right) + C, & \text{(sine with offset)},
\end{cases}
\end{equation}
where $OC(t)$ is the O-C values as a function of time $t$ and all other variables are fit parameters. We restricted our polynomial fit to third degree, as tests with systems exhibiting known TTVs requiring higher-order fits are usually better described by a sine curve of combination of polynomial and sine. The period of the sine curve was limited $3\times$ the planet period. To determine which model best describes the data while penalizing over-fitting, we computed the \textit{Bayesian Information Criterion} (BIC) for each fit,
\begin{equation}
\mathrm{BIC} = k\ln n + \sum_{i=1}^{n}\!\left[\frac{r_i^2}{\sigma_{OC}^2} + \ln(2\pi\sigma_i^2)\right],
\end{equation}
where $k$ is the number of free parameters, $n$ is the number of data points, 
$r_i$ are the residuals, and $\sigma_{OC}$ are the timing uncertainties. While real TTVs are often more complicated than these forms, it is generally the case that a real TTV will be better described by a sine curve or (for long-period TTVs) a polynomial than a linear fit. So a lower BIC for either case is indicative of a TTV.

Prior studies of TTVs from \kepler{} generally use a wider array of tests to search for TTVs, such as periodograms and the correlation between adjacent O-C values \citep[e.g.,][]{Tamuz2006,Mazeh2013}. These are largely impractical for our dataset due to the number of transits and gaps in the coverage. An additional problem is the inhomogeneity of our dataset; we need a metric that works across targets with as few as five transits and as many as 500. For example, \cite{Mazeh2013} uses a metric that compares the median-absolute deviation of the timings to that of the typical uncertainty in the timings. That works well for \kepler, because the data is relatively homogeneous and hence timing uncertainties are similar over the whole dataset. Transit time uncertainties can vary by a factor of $\simeq3$ within our dataset due to mixing of datasets across surveys and changes in the observational cadence over the course of the \tess{} mission.

For this, the $\chi^2_\nu$ is simple and effective. The main problem with $\chi^2_\nu$ is outliers. These often arise because of bad fits and are common when doing bulk analysis of tens or hundreds of thousands of transits \cite{Mazeh2013, Holczer2016}. Fortunately, our sample was small enough to do visual inspection of each, which reduces or eliminates such bad fits and makes $\chi^2_\nu$ a more suitable metric.

\section{Results}\label{sec:results}

\begin{figure*}[ht]
    \centering
    \includegraphics[width=0.45\textwidth]{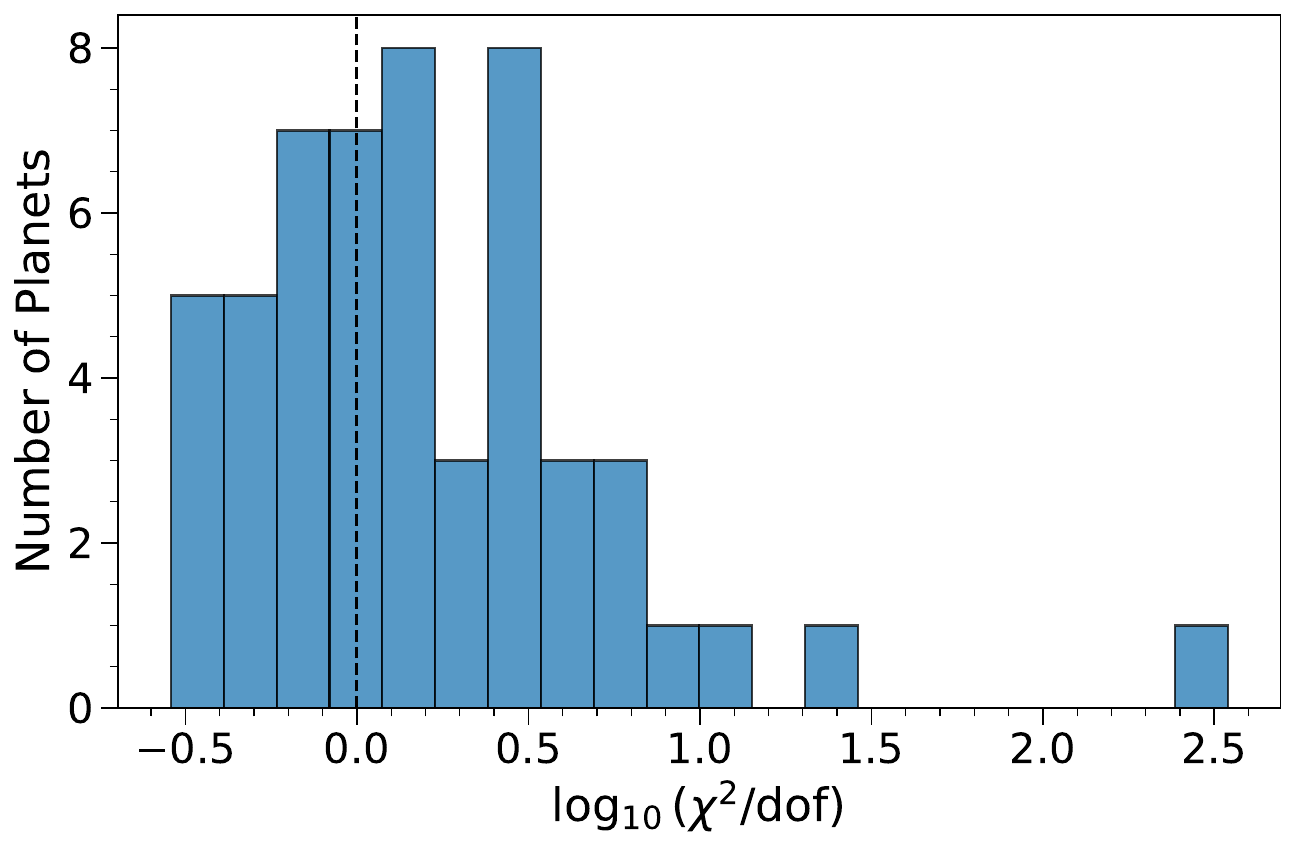}  
    \includegraphics[width=0.45\textwidth]{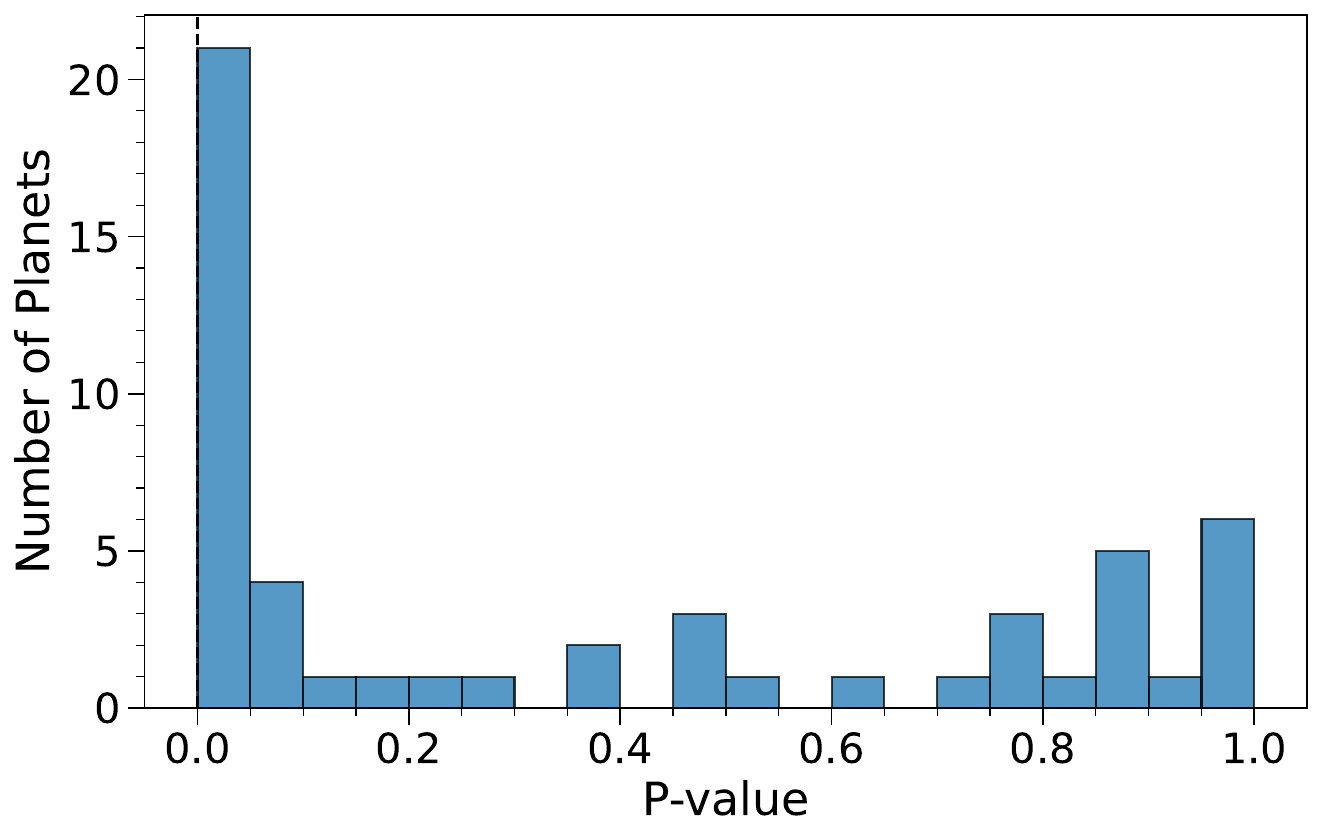}  
    \caption{Histogram of the $\log_{10}{\chi^2_\nu}$ values for all planets (left) and the resulting P-values (right). We marked the left panel with a dashed line at $\chi^2_\nu=1$. The P-values show a peak around zero, indicating likely detections of TTVs. The excess at $\simeq$1 are likely due to overestimated uncertainties.
    \label{fig:histo}}
\end{figure*}

We show the histogram of $\chi^2_\nu$ values in Figure~\ref{fig:histo}. Because there are varying numbers of transits, there is no single $\chi^2$ threshold for significance. Instead, we show the resulting P-values (likelihood of getting $\chi^2$ or larger by chance) in Figure~\ref{fig:histo}. 

To turn the P-values into true probabilities, we need a prior on the underlying probability of a TTV. Normally, this would be small, as only a few percent of systems show detectable TTVs. However, young planets are known to be near orbital resonances at a much higher rate than their older counterparts \citep{Dai2024}. Assuming an equal probability of TTV as no detectable TTV, we identify 16 planets in 13 systems with evidence of significant TTVs ($P<0.3\%$).

For the BIC comparison, 21 planets have fits that favor a polynomial or sine curve over a linear fit. For two of these (HD 110082\,b, and Kepler-1928\,b), the $\Delta$BIC is $<1$, so we remove these. Of the remaining 19, 15 overlap with the targets with significant $\chi^2$. The one planet passing the $\chi^2$ criterion but failing the BIC test is K2-136\,c. The four that pass the BIC test but not the $\chi^2$ test are HD 178085\,b, TOI-1097\,c, TOI-1097\,b, and V1298 Tau\,c. V1298\,Tau is known to show TTVs and TOI-1097bc are within 1\% of a 3:2 period ratio. This suggests the BIC-test is mostly reliable and more sensitive than the $\chi^2$ test. However, we conservatively only consider the 15 that pass both tests as significant candidate TTVs with more data needed to investigate the tentative signals from TOI-1097bc and HD 178085\,b.

\begin{figure*}[ht]
    \centering
    \begin{minipage}{0.48\textwidth}
        \centering
        \includegraphics[width=\linewidth]{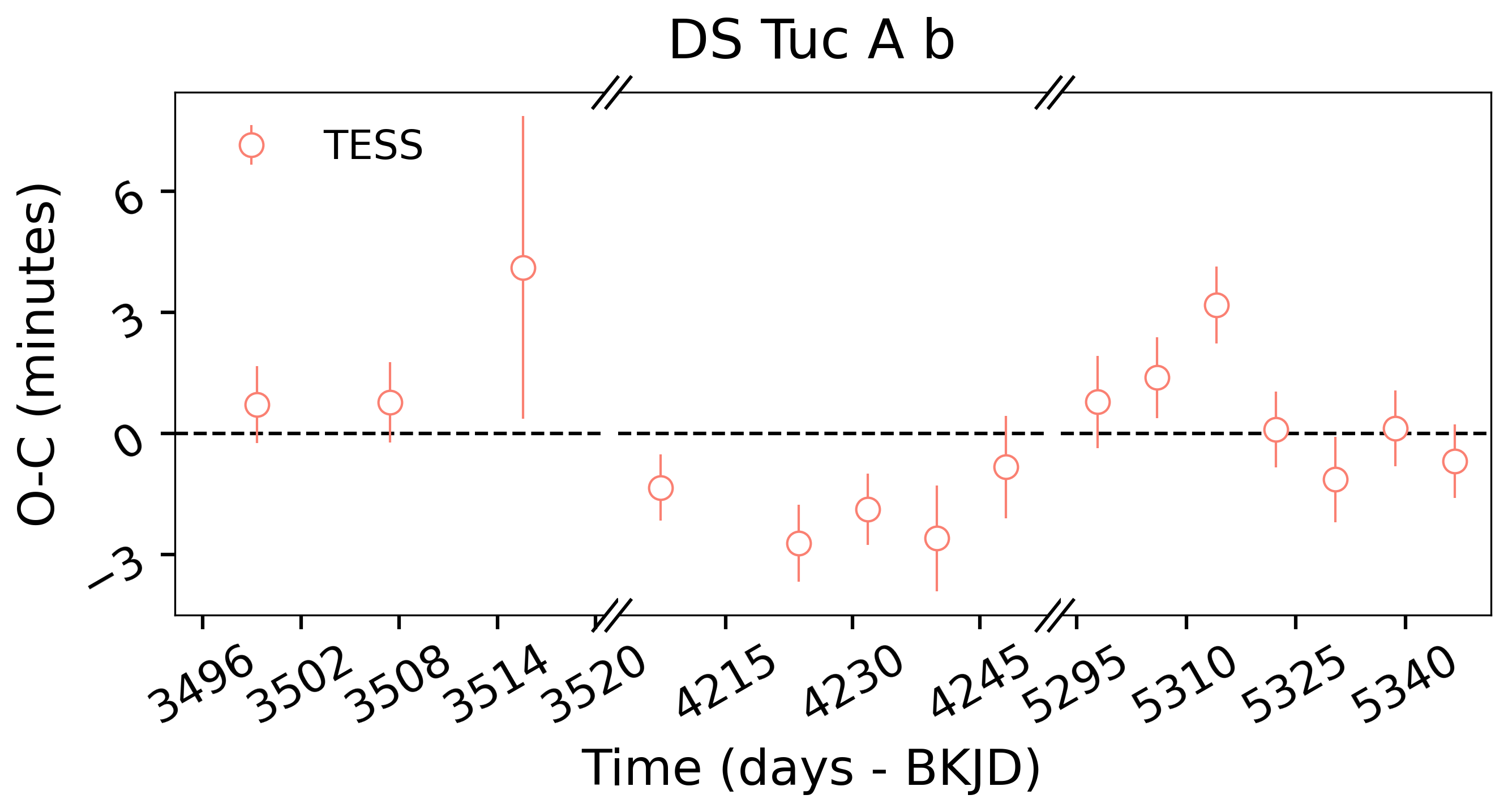}
    \end{minipage}
    \hfill
    \begin{minipage}{0.48\textwidth}
        \centering
        \includegraphics[width=\linewidth]{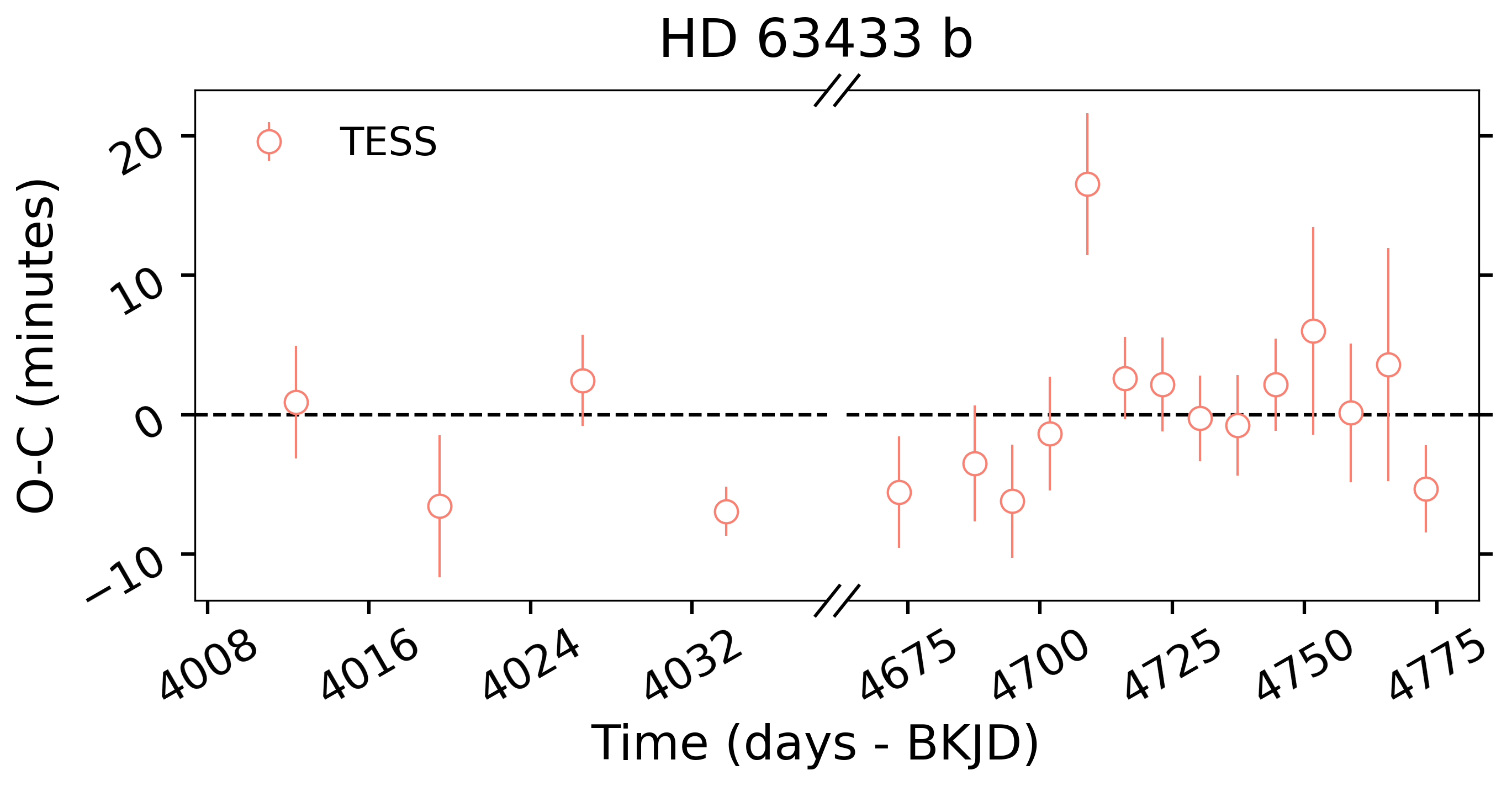}
    \end{minipage}

    \vspace{1em}
    \begin{minipage}{0.48\textwidth}
        \centering
        \includegraphics[width=\linewidth]{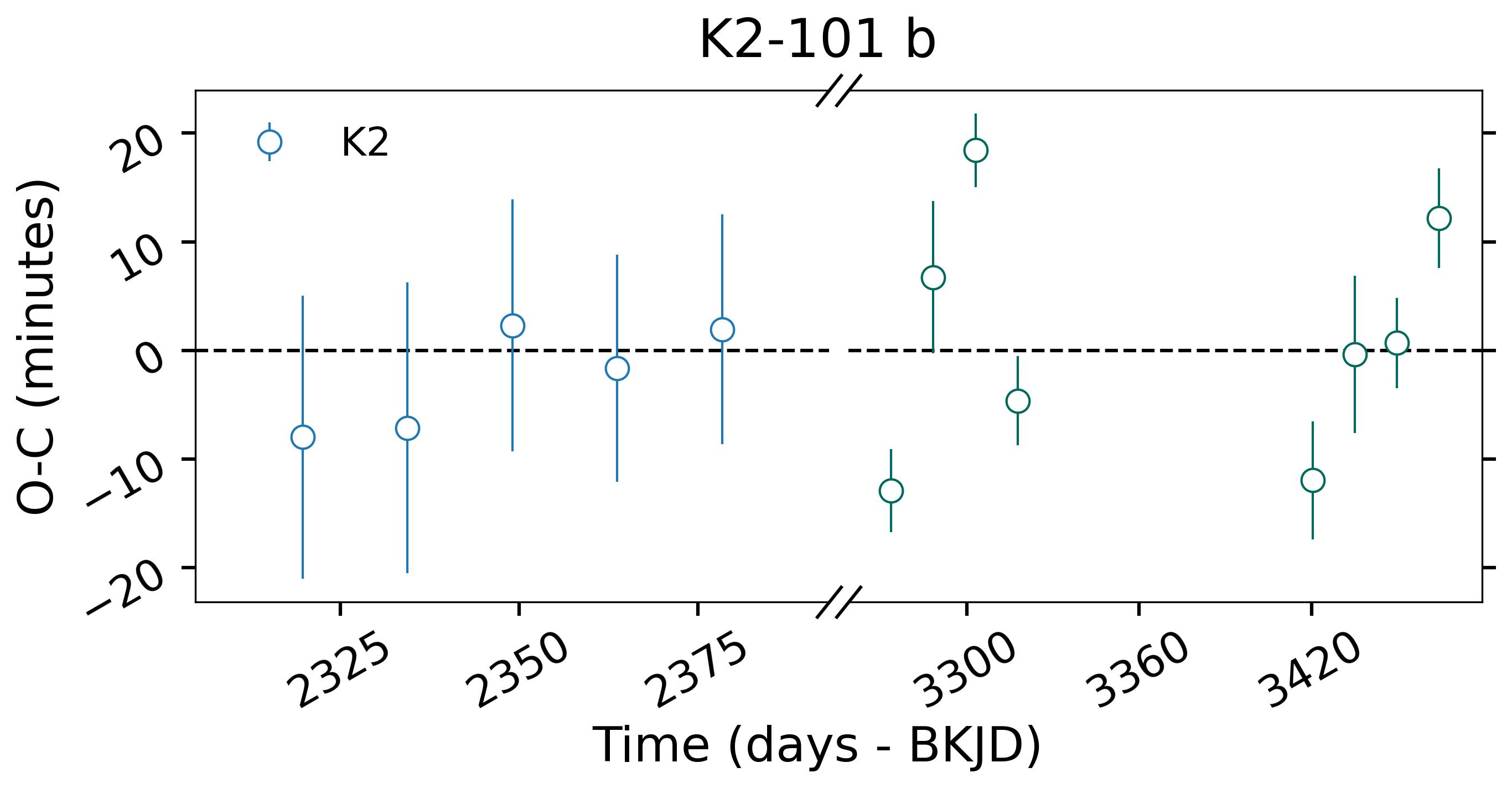}
    \end{minipage}
    \hfill
    \begin{minipage}{0.48\textwidth}
        \centering
        \includegraphics[width=\linewidth]{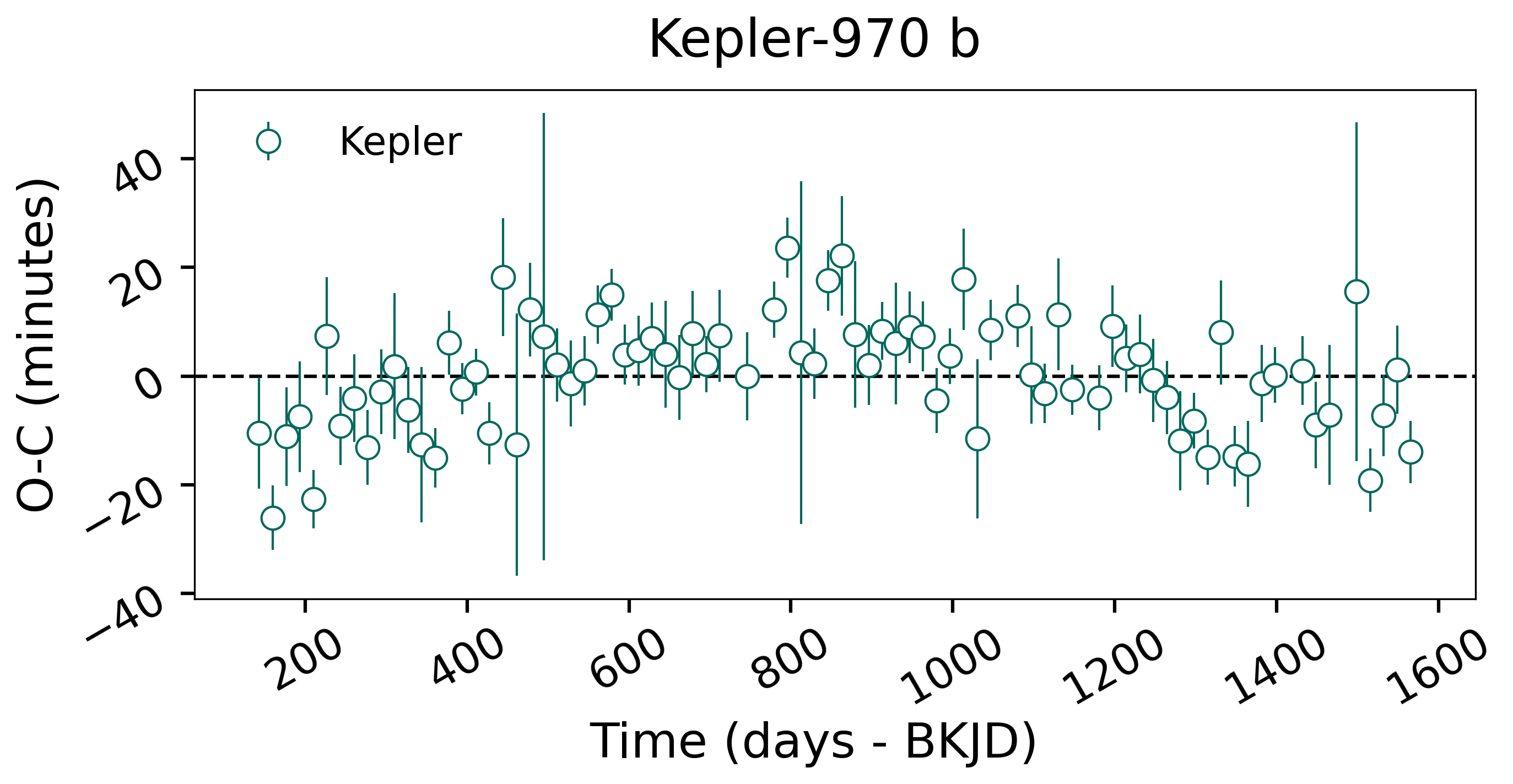}
    \end{minipage}

    \vspace{1em}
    \begin{minipage}{0.48\textwidth}
        \centering
        \includegraphics[width=\linewidth]{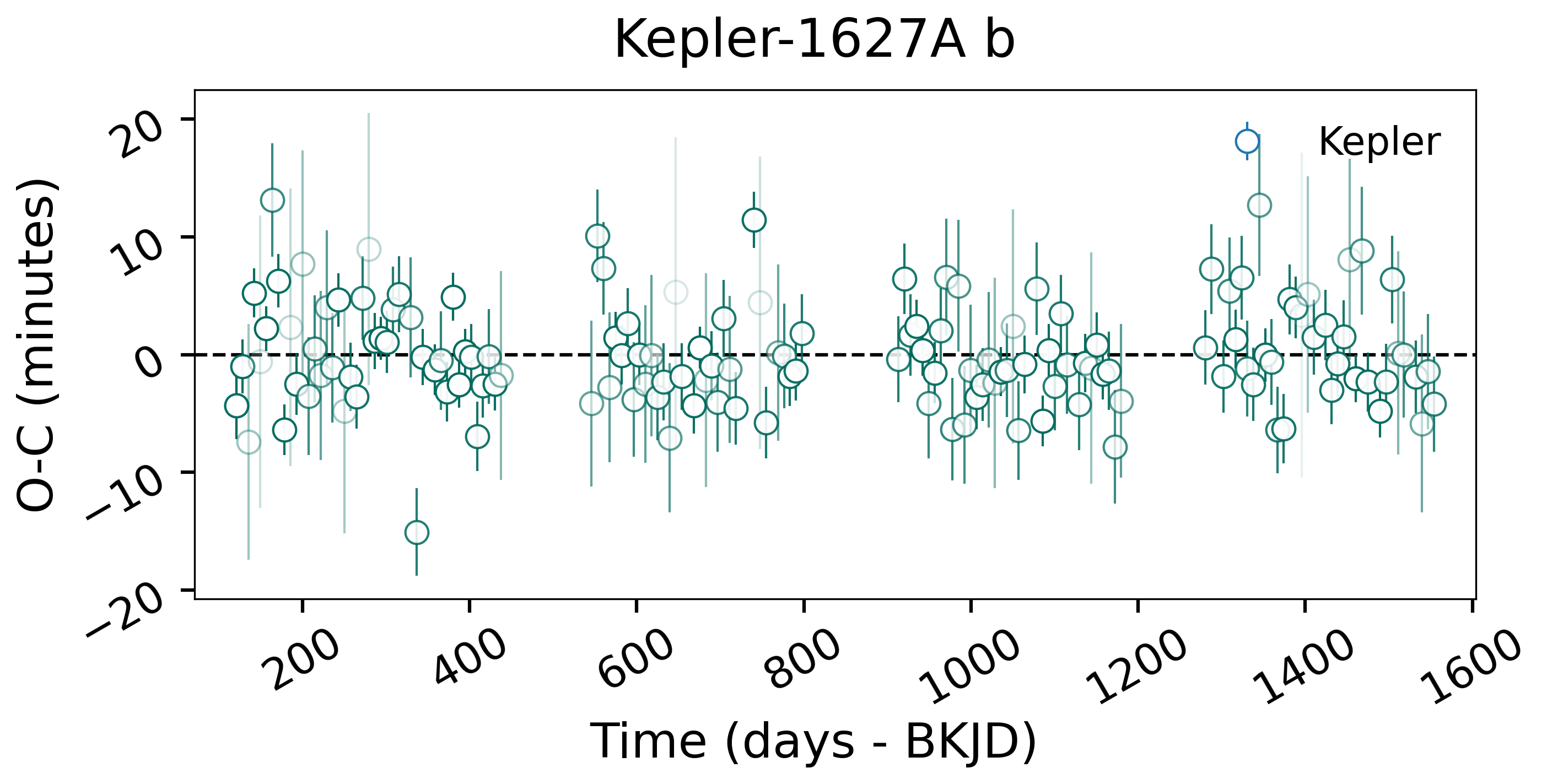}
    \end{minipage}
    \hfill
    \begin{minipage}{0.48\textwidth}
        \centering
        \includegraphics[width=\linewidth]{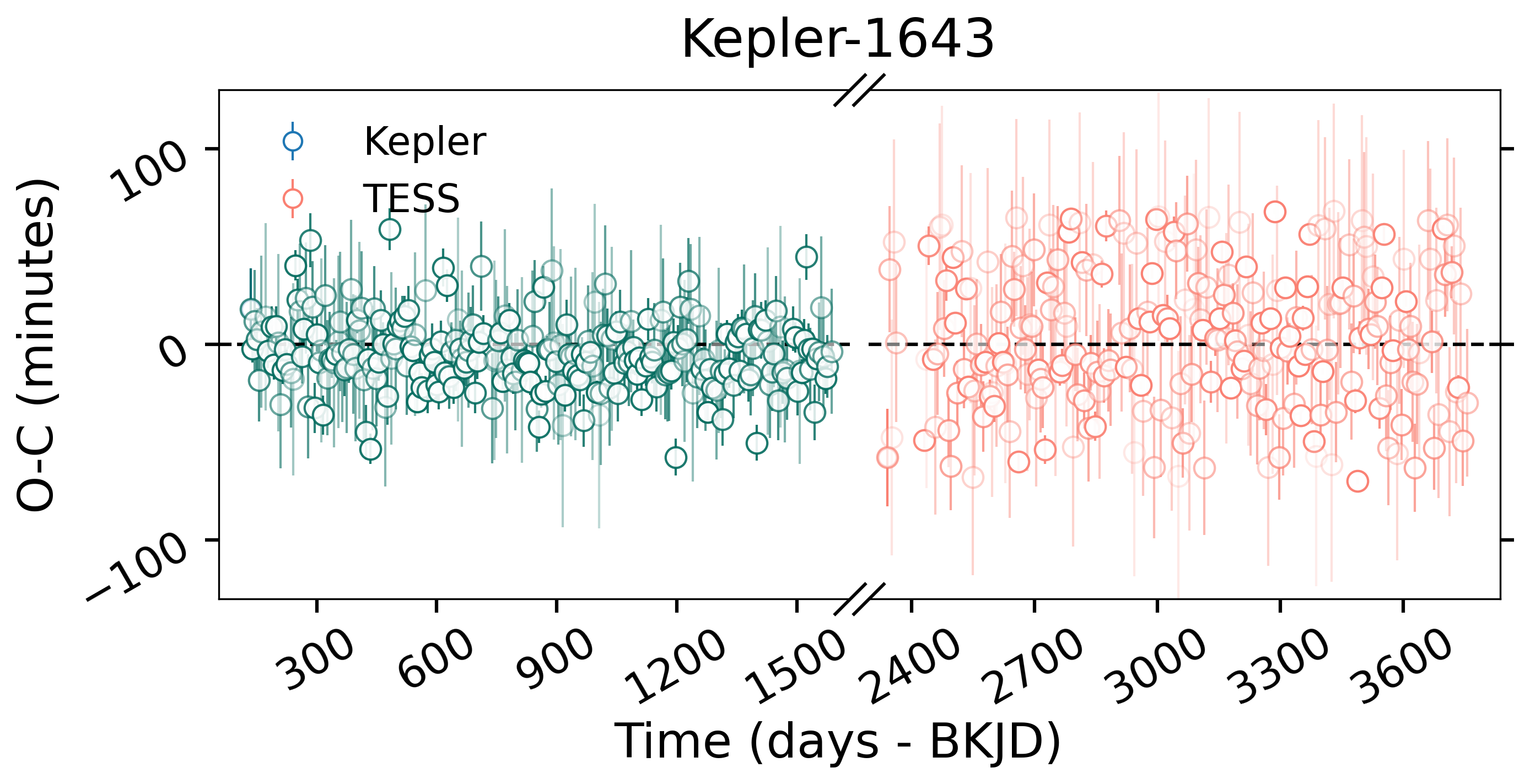}
    \end{minipage}

    \begin{minipage}{0.48\textwidth}
        \centering
        \includegraphics[width=\linewidth]{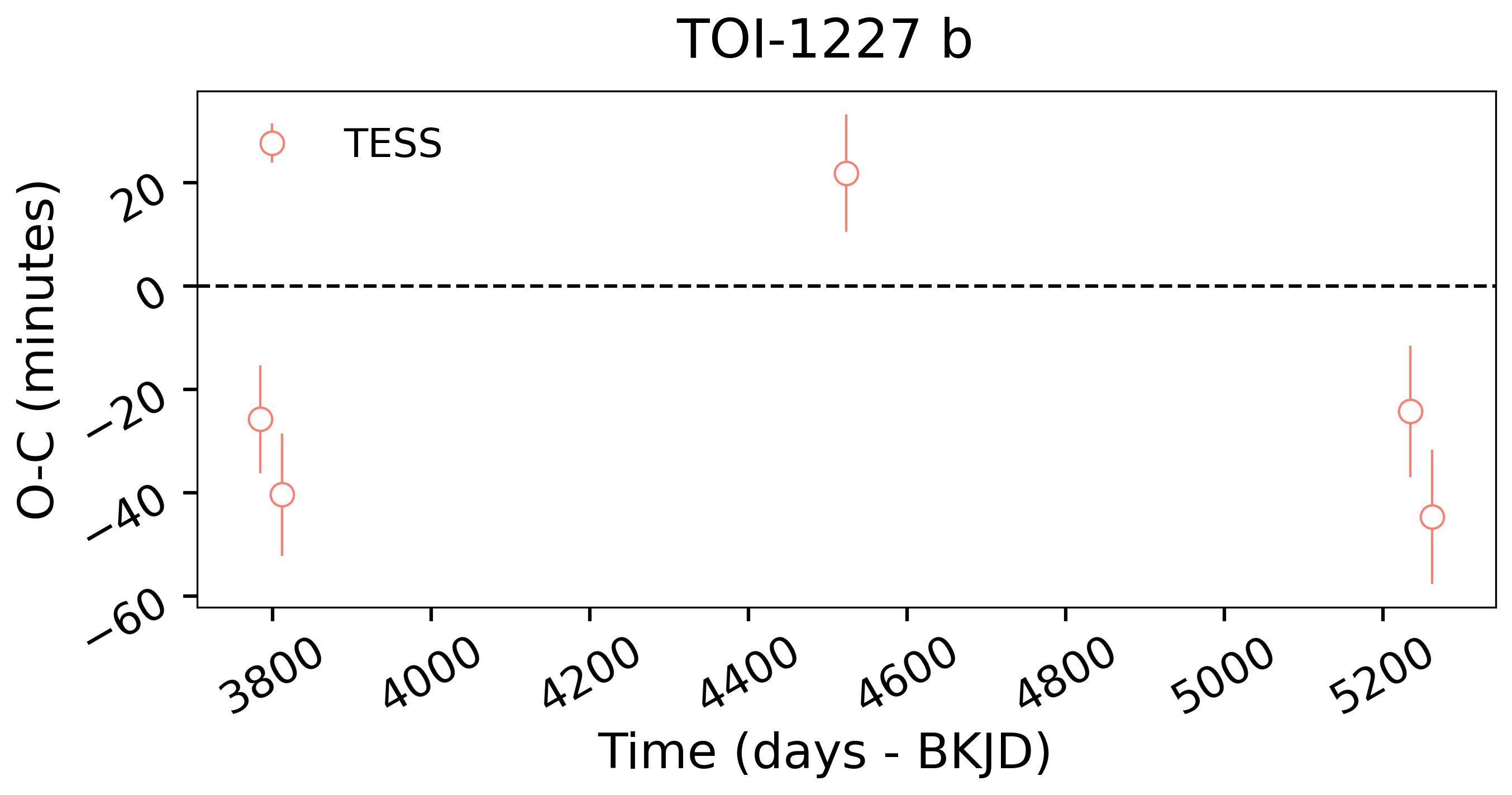}
    \end{minipage}
    \begin{minipage}{0.48\textwidth}
        \centering
        \includegraphics[width=\linewidth]{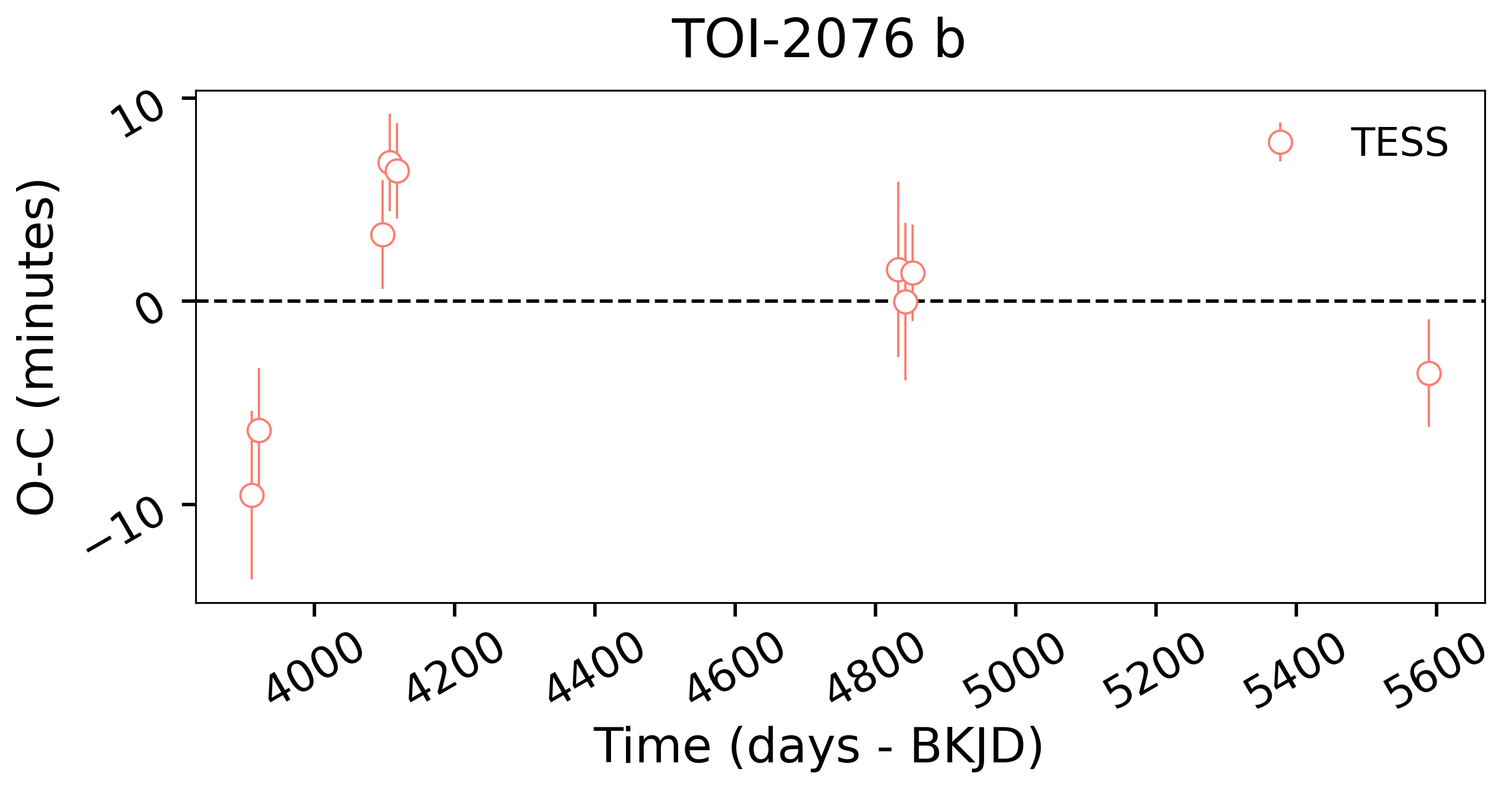}
    \end{minipage}
    \caption{Observed - Calculated (O-C) individual transit times for  planets in our sample with significant TTVs. For planets with high volume of transits (i.e, Kepler-1627 A b and Kepler-1643), O-C transparency has been scaled on uncertainty for visibility. Recent systems with previously reported TTVs similar to our results have been omitted (e.g., AU Mic, Kepler-51) \citep{Wittrock2022, Masuda2014, Masuda12024}.
    \label{fig:oc_plots}}

\end{figure*}

We show the O-C plots of a subset of planets passing both tests in Figure \ref{fig:oc_plots}. Of these, most are in systems known to show TTVs \citep[e.g., Kepler-970, V1298 Tau, and TOI-2076;][]{Holczer2016, Feinstein2022, Almenara2024}. Four of these are new detections: DS Tuc, HD\,63433, K2-101, and Kepler-1643. 

Among the new systems, DS Tuc is the only in a known binary, which raises the possibility that the TTV signal is due to barycenter motion and light travel time of the system (Roemer delay) rather than perturbations from another planet. However, the binary has a projected separation of $\simeq$240 AU, hence a period of (at least) thousands of years. The expected TTV amplitude from the binary is negligible over this window, and it is particularly unlikely that we would see a slope change over so few transits. We conclude that the TTV is likely due to a second planet orbiting DS Tuc A.

Five planets exhibiting TTVs are in systems with a single known transiting planet (DS~Tuc\,A\,b, K2-101\,b, Kepler-1627\,Ab, Kepler-1643\,b, and Kepler-970\,b). As with DS Tuc (discussed above), these likely harbor another (undetected) planet. Kepler-970\,b's TTV was initially attributed to an unseen companion \citep{Barber2022}, mostly based on the long TTV period ($\gtrsim$1000\,days). However, no companion is visible in the \gaia\ imaging, the target has a RUWE of 0.98 \citep[RUWE of $>1.4$ is usually indicative of binarity;][]{Wood2021}, and the star sits tightly on the CMD expected for its age. While we cannot rule out a low-mass companion, it seems more likely this is due to a non-transiting outer planet. 

Transit times are especially challenging for Kepler-1643 because of the shallow depth of the transit, making it less visible in the \tess\ data. Despite that, it shows significant TTVs in both datasets, with a large maximum TTV amplitude of over 50 minutes. This system had the largest amount of transit sampling ($490$ transits over the combined \kepler\ and \tess\ data) in our target selection. Unfortunately a full solution for the mass and eccentricity is unlikely unless the perturber also transits and is recovered.

Kepler-1627\,Ab's TTV was discussed in more detail in \citet{Bouma2022}. \citet{Holczer2016} found no evidence of a long-term TTV, although they see significant (short-term) TTVs. Indeed, their O-C times are often further from zero than ours, with many transits off from the linear ephemeris by 20-50m (typical uncertainties are 3-5m). \citet{Bouma2022} revisited this and found O-C values closer to ours, and re-affirm the lack of {\it long-term} TTVs. They also find weak/suggestive evidence that the short-term TTVs are caused by spots based on the out-of-transit slope \citep{Mazeh2015}, as well as evidence of asymmetry in the transit (which might or might not be related to the TTV). We see no evidence that the TTVs are spot-induced and consider this TTV significant, although the TTV signal itself does not show the characteristic long-term pattern seen in systems like Kepler-51.

\begin{figure}[ht]
    \centering
    \includegraphics[width=0.48\textwidth]{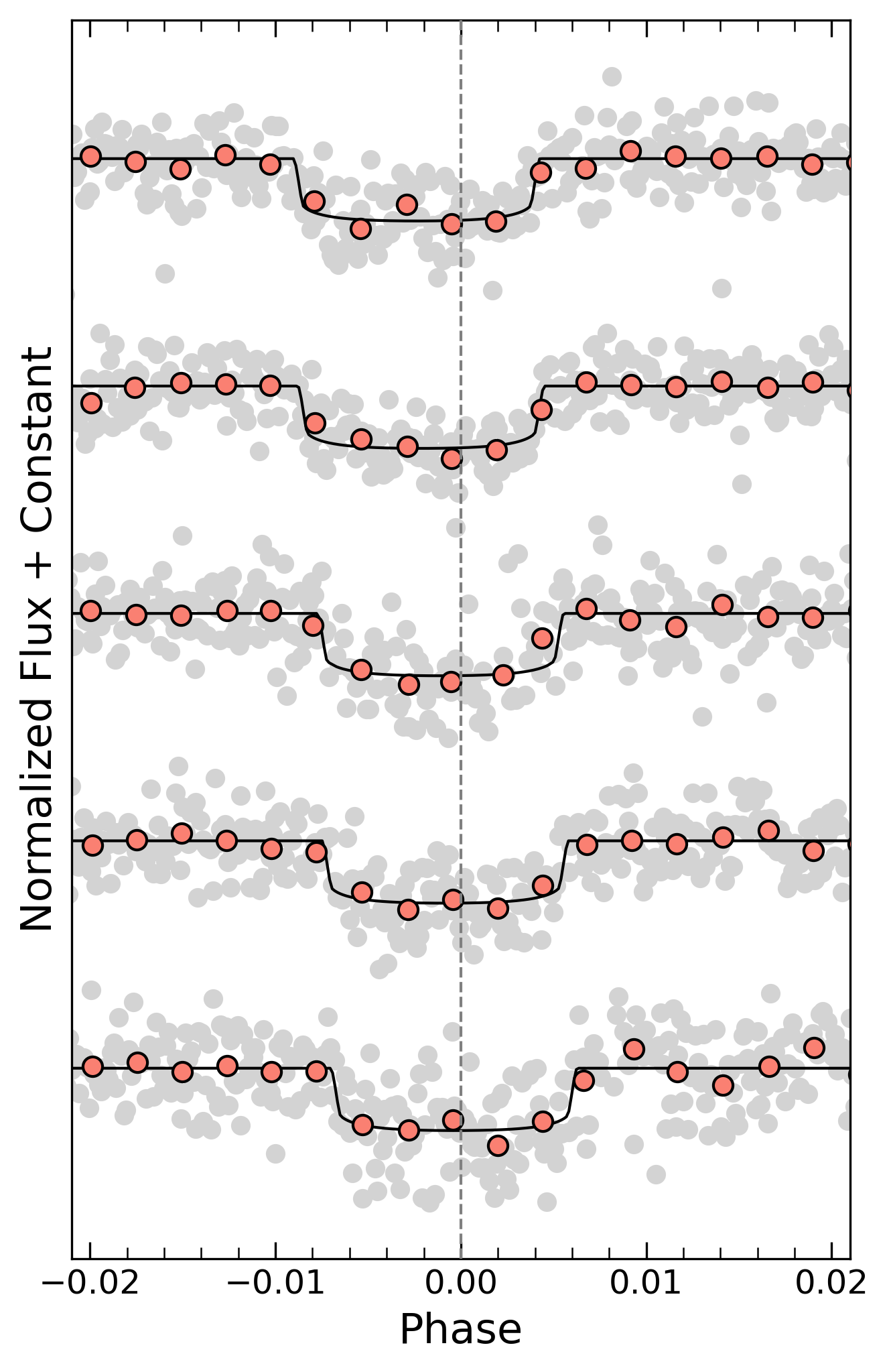} \caption{Individual transit of TOI-2076 b transits taken with \tess, folded to the linear ephemeris. Our best-fit model is overlaid in a black line. Circular dots represent data binned into 3.5 minute intervals. 
    \label{fig:2076_visualization}}
\end{figure}

The V1298 Tau system is the subject of a more detailed TTV analysis to measure its mass with significantly more transit data than analyzed here (Livingston et al. accepted). It is noteworthy, however, how strong the TTV signal is from \ktwo\ and \tess\ alone, such that the timings are visible by eye (Figure~\ref{fig:oc_plots}). The TOI-2076 system has also been studied for mass measurement constraint due to its periodic TTV and amplitude of $\sim$ 10 mins \citep{Osborn2022}. We show an example visualization plot of TOI-2076 b's TTV in Figure~\ref{fig:2076_visualization}.

We see an excess of systems with low $\chi^2$ values (P-values near 1); Figure~\ref{fig:histo}b should show a near-uniform distribution with a peak near $\simeq$0 (due to real detections). We attribute this to overestimated uncertainties due to overfitting by the GP. Some of these systems show unusually broad $T_0$ posteriors given the data precision. This can happen for weaker transits and/or those near data gaps or flares; the GP opts to weaken or distort the transit in order to explain away other oddities in the data. While we attempt to manually flag obvious/large flares and remove some partial transits that are challenging to fit, more subtle effects can be hard to distinguish by eye and we wanted to keep our fitting method as homogeneous as possible. 

Fortunately, the number of systems with unexpected large timing errors is small; there are 1--3 planets more than expected for a uniform distribution. Further, the only multi-planet system in this group was K2-233, which does not have period ratios that would generally yield detectable TTVs. Slightly overestimated uncertainties may cause some systems with marginal signals (like TOI-1097bc) to be below our $\chi^2$ requirements, although the BIC test would catch them.

\begin{figure*}[ht]
    \centering
    \includegraphics[width=0.95\textwidth]{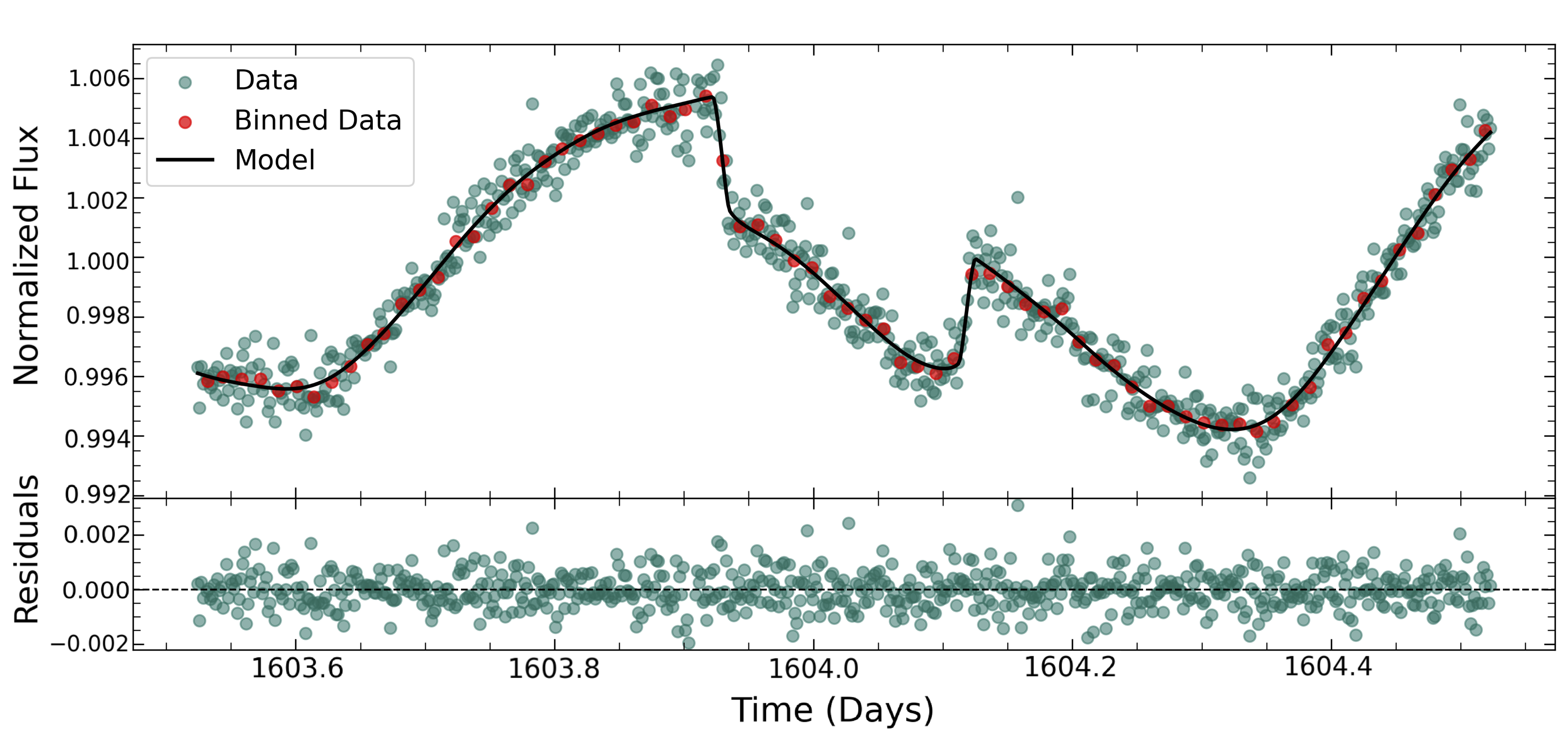}  
    \caption{The first transit of HIP~67522\,b (green points) and the best-fit model (black). For clarity, we also show the data binned to 20 minutes (red points). The bottom panel shows the best-fit residuals, which are generally consistent with white noise and suggest the GP is handling the strong stellar variability well in most cases. 
    \label{fig:hip67522}}
\end{figure*}

Visual inspection of each fit also confirmed that such cases of overfitting are not due to strong rotational modulation (which would increase the bias towards inactive/older systems). HIP 67522, for example, is one of the youngest targets, with some of the strongest variability and the shortest rotation period \citep[1.4\, days,][]{Heitzmann2021}, which led to large-scale modulations even in narrow windows around the transit. As we show in Figure~\ref{fig:hip67522}, the GP handled the complex stellar variability quite well.

\subsection{Can spot activity explain some TTVs?}\label{sec:spots}

One potential explanation for the high TTV rate among young planets is spots/activity. Young stars often exhibit fast rotation \citep{Cody2017} and significant starspot coverage \citep{Morris2020, Barber2023} when compared to their older counterparts. If the planet crosses a spot, especially near ingress or egress, it can introduce a TTV due to changes in the shape of the transit. 

To test this effect, we injected 3000 randomly generated spots into a transit pulled from our fit sample (DS Tuc Ab). We use the spot parameters from \cite{Thao2024_featherweight}, which detected spot crossings in \jwst\, data during the transit of HIP\,67522b, and the analysis in \cite{Morris2017} as a guide for our injected spot parameters. We vary the spot amplitude (0.01-0.1\%) and duration (2-15 minutes) and intentionally place the spot mid-point during ingress or egress to maximize the effect. We then refit the transit using the same framework as above. 

\begin{figure*}
    \includegraphics[width=.63\textwidth]{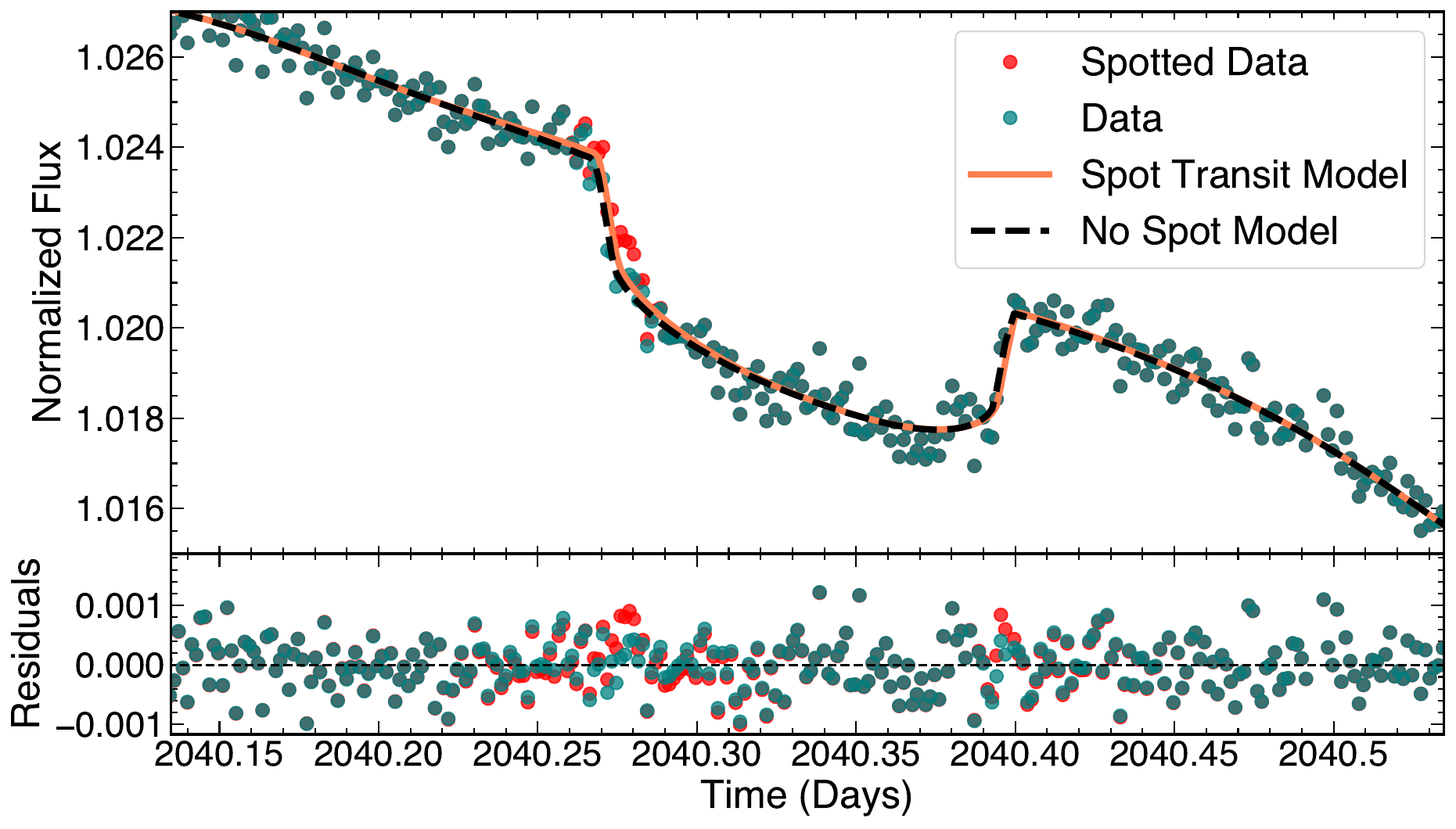}
    \includegraphics[width=.37\textwidth]{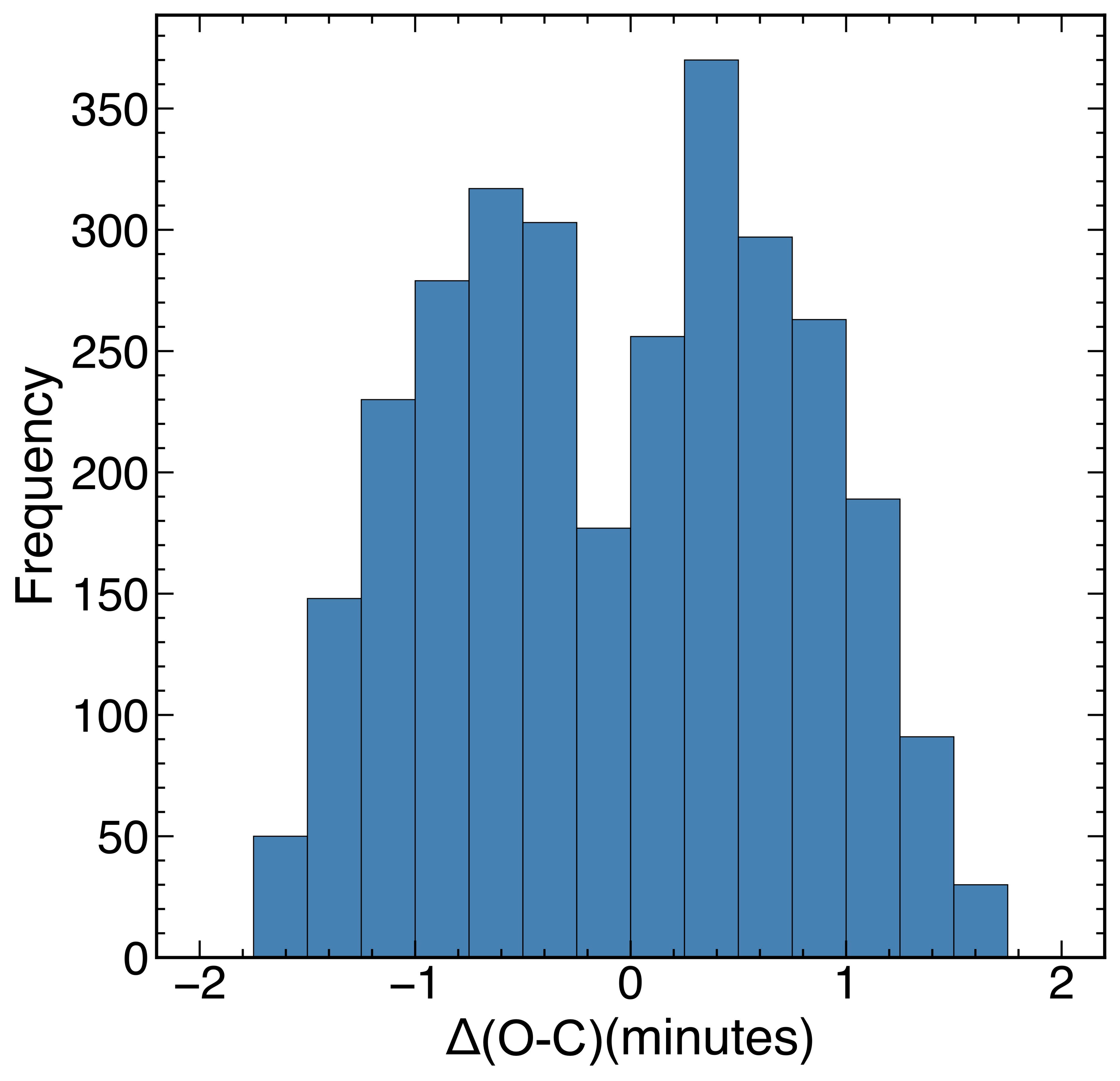}
    \caption{Left: Transit of DS Tuc Ab (teal) with an example injected spot (red). The orange line shows the model the spot-injected transit and the black line shows the model of the original transit. The residuals for each fit are shown in the bottom panel. Right: Distribution of the change in calculated O-C once the spot was introduced. All O-C results from the spot-injected transits were within 2 minutes of the assumed no-spot transit. \label{fig:spot}}
\end{figure*}

In all cases, the impact on $T_0$ was small ($<2$\,minutes;  Figure~\ref{fig:spot}), and well below the measurement uncertainties. Inspection of the fits revealed the reason; spot crossings large enough to induce a significant offset in the transit time also show up as an asymmetry that cannot be explained within the \texttt{BATMAN} model, and hence are modeled out by the GP. The same effect was seen for spot-crossings in HIP\,67522\,c \citep{Barber_HIPc}; the GP generally fits these out by enforcing a symmetric transit model. In such cases, the more important impact was an increase in the measurement uncertainties on $T_0$ to account for the stronger GP and transit asymmetry. However, we did not need to tell the GP there was a spot for this to happen; hence, the larger uncertainties are already included in our fit posteriors. 

If spots were responsible for the TTVs detected here, this would show up as a correlation between spot coverage and TTV amplitude. To check this, we approximate variability amplitude using the normalized 90th percentile of the amplitude of the light curve minus the 10th percentile \citep{Boyle2025}. The connection between variability and spot coverage is complex \citep{Luger2021a,Luger2021b}, but this parameter is at least correlated to spot coverage \citep{Morris2020, Barber2023}. Based on this, we find that systems with detected TTVs are {\it less} active than their counterparts (show a smaller variability amplitude). Indeed, we did not detect TTVs in some of the most active stars in the sample (e.g., HIP\,67522 and IRAS 04125+2902). Some of this is observational bias; it is easier to robustly detect the TTV for well-behaved light curves, and hotter/brighter stars tend to have fewer spots. HIP\,67522 has a previously confirmed TTV signal \citep{Thao2024_featherweight}, suggesting that if there is an effect, spots are causing us to overestimate our $T_0$ uncertainties and hence miss real TTVs rather than activity generating non-planetary TTVs. 

Lastly, \citet{Mazeh2015} suggested that for a planet on a prograde orbit, spot-induced TTVs will be correlated to the out-of-transit variability. This assumes a single (dominant) spot cluster is responsible for both the stellar variation and the induced TTV from the planet crossing the spot, in which case, positive TTVs will show up preferentially when the stellar flux is decreasing (negative slope). We show the result of this test in Figure~\ref{fig:spot_slope} for a representative set of systems. We see no evidence of a correlation, including for Kepler-1627A\,b, providing further confirmation that spots have a negligible or minor impact on transit times. 

\subsection{TTV occurrence for young versus old systems}

Of the 53 young planets considered here, 15 pass both TTV tests show significant TTVs ($28.3\pm6.2\%$). This is significantly higher than the TTV rate seen among systems harboring Kepler Objects of Interest (KOIs). \citet{Mazeh2013} searched 1960 KOIs and found 143 of them $(7.3\pm0.6\%)$ showed significant TTVs. Similar studies have found comparably low rates. \citet{Xie2014} estimate that the TTV fraction is only a few \% for single-planet-transiting systems, rising to 10\% for 4-planet systems. The full \kepler\ sample yielded a rate closer to 10\% \citep{Holczer2016}. This is marginally consistent ($\lesssim3\sigma$) with our TTV fraction, although as we discuss below, \kepler{} has a much longer baseline to detect TTVs. More realistic simulations of TTVs with \tess\ data, such as those from \citet{Ballard2019}, estimate only $\simeq$5\% of planets from \tess\ should exhibit TTVs and the fraction of planets with {\it detectable} TTVs will be below that \citet{Hadden2019}, placing it $\ge4\sigma$ below that of our TTV fraction. 

\kepler\ stared at a single field for $\simeq$4.5 years, which provides many more transits than \tess's strategy of $\simeq$27-day sectors with multi-year gaps between. \citet{Holczer2016} had, on average, 113 transits per KOI, while our sample has an average of 41 transits per planet (median of 13), a factor of 3 (9) difference. The only young systems with comparable numbers of transits ($\sim$hundreds) are those in the \kepler\ field (Table~\ref{tab:obs}). To explore the effect, we simulate \tess-like coverage using the transit times from \citet{Mazeh2013} and \citet{Holczer2016}, randomly sampling coverage taken from our sample (Table~\ref{tab:obs}), but keeping the timing uncertainties. Since our sample includes \kepler\ targets, some targets will (randomly) retain their original sampling. In this situation, 35\%--55\% of their TTVs become insignificant. The resulting fraction of systems with TTVs assuming similar data is $3.2^{+0.8}_{-0.6}$\%, $4\sigma$ below our detection rate. This fraction is also slightly elevated because it counts some young systems.

\kepler\ systems tend to have 30-minute cadence while \tess\ systems considered here mostly have data taken at 10-minute or faster. Indeed, we had to exclude some 30m \tess\ data because it was too challenging to detect the TTVs. However, our need for the more rapid cadence is largely driven by our method for fitting stellar variability; the signal is significant over the transit duration (especially compared to older \kepler\ targets) and there are typically too many free parameters to fit transit and variability parameters simultaneously when the transit contains too few points. Nearly all our detections would be significant with 30-minute cadence absent stellar variability (based on formal errors). Further, a comparison of transit times from our analysis and prior ones suggests our GP is overestimating our timing uncertainties and hence underestimating the fraction of systems showing significant TTVs. We conclude that the cadence bias favoring TTV detection in young systems is small compared by the bias against TTV detection in these systems due to higher variability. 

Another way to consider this is to look just at those with \kepler\ data. There are seven systems, four of which harbor planets with candidate TTVs (50\%): Kepler-51bcd, Kepler-970b, Kepler-1627Ab, and Kepler-1643b. The low end of the $3\sigma$ (99.7\%) confidence interval gives a TTV fraction of 15\%, which is still outside the range for the \kepler\ sample (5.6--9.2\% at 99.7\%). If we exclude Kepler-1627Ab (see Section~\ref{sec:spots}) the TTV fraction is $>$9.6\% (99.7\%), still inconsistent with the total \kepler\ TTV sample. 

\begin{figure*}[ht]
    \centering
    \includegraphics[width=.99\textwidth]{combined_spot_plot.png}
    \caption{Left: TTV residual (O-C) in minutes as a function of the out-of-transit slope (the flux derivative) for five (representative) planets showing significant TTVs. Right: the same for Kepler-1627A\,b as well as the best-fit line (red) and uncertainty (shaded region). In the case of spot-induced TTVs, a prograde orbit, and a single dominant spot, these parameters should be correlated \citep{Mazeh2015}. No planet we tested shows a significant correlation; it is unlikely that spot crossings are a significant source of TTVs. 
    \label{fig:spot_slope}}
\end{figure*}

The last complication for this comparison is the metric used. As noted in Section~\ref{sec:results}, \kepler\ studies identify TTVs based on an additional set of metrics, including the scatter compared to the typical uncertainty, consistency between consecutive transits, and evidence of periodicity in the data \citep{Mazeh2013,Holczer2016}. The dataset here is too narrow to resolve this difference, but we discuss options for future analysis in Section~\ref{sec:conclusion}.

\section{Conclusions and Discussion}\label{sec:conclusion}

We have analyzed the transit light curves of 53 young planets detected by \kepler, \ktwo, and \tess. We have leveraged data from all three surveys, fitting a total of 1922 individual transits over a baseline of up to 7 years. We used the individual transit fits to search for transit timing variations in these systems, with the goal of identifying systems that are amenable to follow-up to characterize their masses and eccentricities. 

We found evidence of TTVs for 15 planets in 12 systems. Four of these are newly reported candidate TTVs, including the three-planet system HD\,63433 \citep{Capistrant2024}. An additional four planets show weak evidence of TTVs (passing one of our tests). This includes TOI-1097bc (two planets near 3:2 resonance) and V1298 Tau\,c (which has a known TTV). These muftis are the most amenable for additional follow-up to fully characterize the TTV. 

Our survey did not include or missed some young systems with known TTVs, notably HIP\,67522bc \citep{Thao2024_featherweight}, AU Mic\,c \citep{Wittrock2022}, and TOI-2076cd \citep{Osborn2022} both had too few transits in \tess{} alone for us to fit. HIP\,67522 had more transits, 10 for b and 5 for c, but the TTV has a low amplitude (5-20\,minutes), and was only clear with \jwst\ and ground-based data analyzed alongside the \tess\ data \citep{Thao2024_featherweight}. This is consistent with suggests that many more systems will have detectable TTVs when the \tess\ mission revisits these targets. 

We find that young planets may exhibit TTVs at $>$3 times the rate of their older counterparts (from \kepler). Differences in the number of transits increases this difference. A completely fair comparison remains challenging due to differences in data quality as well as the smaller number of young planets and number of transit per planet. Repeating an identical analysis on the older \kepler\ sample is also impractical due the heavy computational cost of the GPs. However, the suite of tests we run is strongly suggestive of a significant difference between TTV rates of young and old systems. Further evidence of this is that many young planets are known to have TTVs but were not detected here due to the small number of transits (see above); they would have been easily detected with \kepler-like data coverage. The higher rate of TTVs is also consistent with prior findings that young planets are more likely to be found near period resonances when compared to their older counterparts \citep{Dai2024}.

A more robust approach to exploring variations in the TTV rate with age would be a comparison {\it between} young planets. As resonances break up, the rate of TTVs should go down and this is likely to be detectable over the first Gyr. A homogeneous analysis of 0-50\,Myr, 50-200\,Myr, and 200-1000\,Myr systems could elucidate how resonant orbits form and break up and reduce issues when comparing to samples that use different methods and with completely different coverage/data. Such a test would require roughly 30-50\% growth in the population of young transiting planets with known ages. This seems high, but \tess\ data has been already used to reveal additional planets since we adopted our sample \citep[e.g.][]{Distler2025,Vach2025,Barber2025}. The science could be achieved with much more modest growth in the sample when combined with longer baselines expected from further extensions of the \tess\ mission, as well as better age-dating of known transiting systems \citep[e.g.,][]{Barber2023,Bouma2023}.

Spots have a weak (or no) impact on our results. Experiments injecting transits and checking for correlations between TTVs and out-of-transit slope suggest spot crossings increase the uncertainties, but do not bias the results, nor do they induce TTVs. This indicates that the difference between young and old systems cannot be explained by differences in spot coverage or activity levels. If anything, this suggests the opposite; the larger uncertainties arising from spot crossings and stellar variability mask otherwise detectable transit timing variations. 

The high rate of TTVs, the weak impact of spots on TTVs, and the challenges of measuring the masses of young planets with radial velocities \citep[e.g.,][]{Blunt2023} combine to make a strong case that TTVs are a strong route to measure the masses and eccentricities of young planetary systems, particularly when combined with more sophisticated modeling including photodynamical modeling \citep{Mills2016}. Indeed, the only other successful method for measuring masses of $<50$\,Myr planets has been via atmospheric scale height \citep{Thao2024_featherweight,Barat2024} which is impractical for large numbers of systems and is likely to fail on systems with significant hazes or cloud-cover \citep[e.g.][]{Thao2023}. This is limited to systems with multiple detected transits, but that includes 9 of the 15 planets with TTVs in this work. Many of the single-transiting systems with TTVs may also have transiting planets that have not yet been detected; the recovery and validation of HIP\,67522c, for example, began because of the TTV detection in HIP\,67522b \citep{Thao2024_featherweight,Barber_HIPc}.

\acknowledgements

We thank the anonymous referee for their careful reading and thoughtful comments on the manuscript. 
The authors also wish to thank Bandit and Halee for their detailed discussions on TTV posteriors and pet treats. AILM was supported by the NC Space Grant Graduate Research Fellowship (80NSSC25M7117) and the Chancellor Science Scholars\footnote{\url{https://chancellorssciencescholars.unc.edu/}} programs. AWM was supported by grants from the NSF CAREER program (AST-2143763) and NASA's exoplanet research program (XRP 80NSSC25K7148). MGB was supported by NSF Graduate Research Fellowship (DGE-2040435). PCT and AILM were supported by a grant from NASA's \tess\ Guest Investigator program (80NSSC24K1142). 

We provide access to a GitHub repository including code created for the analysis of this project that is not already publicly available: \hyperlink{https://github.com/AnaLopezMurillo/PIVOT}{https://github.com/AnaLopezMurillo/PIVOT}.

This paper includes data collected by the TESS mission. Funding for the TESS mission is provided by the NASA's Science Mission Directorate.

This paper includes data collected by the Kepler mission and obtained from the MAST data archive at the Space Telescope Science Institute (STScI). Funding for the Kepler mission is provided by the NASA Science Mission Directorate. STScI is operated by the Association of Universities for Research in Astronomy, Inc., under NASA contract NAS 5–26555.

\software{ \texttt{BATMAN}, \texttt{stella}, \texttt{Lightkurve}, \texttt{Juliet}, \texttt{Celerite}
          }

\facilities{TESS, Kepler, K2}
\clearpage

\bibliography{mannbib.bib, lopezmurillobib.bib}

\end{document}